\documentclass[twocolumn,showpacs,preprintnumbers,amsmath,amssymb]{revtex4}

\usepackage{graphicx}
\usepackage{dcolumn}
\usepackage{bm}

\newcommand{\be}{\begin{equation}}
\newcommand{\ee}{\end{equation}}
\newcommand{\ba}{\begin{eqnarray}}
\newcommand{\ea}{\end{eqnarray}}
\newcommand{\sba}{\begin{subequations}}
\newcommand{\sea}{\end{subequations}}
\newcommand{\barr}{\begin{array}}
\newcommand{\earr}{\end{array}}

\def\A {{{\cal A}}}
\def\B {{{\cal B}}}

\def\E {{{\cal E}}}

\def\X {{{\cal X}}}
\def\Y {{{\cal Y}}}

\begin{document}


\title{Self--Consistent Random Phase Approximation \\ Application to the Hubbard Model for finite number of sites}

\author{Mohsen Jema\"i}
\email{jemai@ipno.in2p3.fr}
\affiliation{  Institut de Physique Nucl\'eaire d'Orsay, Universit\'e Paris-Sud, CNRS--IN2P3 \\
               15, Rue Georges Clemenceau, 91406 Orsay Cedex, France.}
\altaffiliation[Also at the]{ D\'epartement de Physique, Facult\'e des Sciences de Tunis, 
                          Universit\'e de Tunis El-Manar 
                          2092 El-Manar, Tunis, Tunisie. }	       
\author{Peter Schuck}
 \email{schuck@ipno.in2p3.fr}
 \affiliation{ Institut de Physique Nucl\'eaire d'Orsay, Universit\'e Paris-Sud, CNRS--IN2P3 \\
               15, Rue Georges Clemenceau, 91406 Orsay Cedex, France.\\}
\affiliation{ Laboratoire de Physique et Modélisation des Milieux Condensés (LPMMC) - (UMR 5493)\\
              Maison Jean Perrin, 25 avenue des Martyrs BP 166, 38042 Grenoble cedex 9 }

\author{Jorge Dukelsky}
 \email{dukelsky@iem.cfmac.csic.es}
 \affiliation{ Instituto de Estructura de la Materia, Consejo Superior de Investigaciones Cientificas, \\
               Serrano 123, 28006 Madrid, Spain }
\author{Raouf Bennaceur}
 \email{raouf.bennaceur@inrst.rnrt.tn}
 \affiliation{ D\'epartement de Physique, Facult\'e des Sciences de Tunis, 
               Universit\'e de Tunis El-Manar 
               2092 El-Manar, Tunis, Tunisie }

\date{\today}

\begin{abstract}
Within the $1D$ Hubbard model linear closed chains with various numbers of sites are considered in Self Consistent Random Phase Approximation (SCRPA). Excellent results with a minimal numerical effort are obtained for $2+4n$ sites cases, confirming earlier results with this theory for other models. However, the $4n$ sites cases need further considerations. SCRPA solves the two sites problem exactly. It therefore contains the two electrons and high density Fermi gas limits correctly.
\end{abstract}

\maketitle

\section{\label{Intro}Introduction}

Standard Random Phase Approximation (s-RPA) is one of the most popular many body approches known. It was invented in condensed matter physics (see e.g. \cite{1}) and has subsequently spread to almost all branches of physics, including atomic physics \cite{2}, molecular physics \cite{3}, plasma physics \cite{4}, relativistic field theory \cite{5}, nuclear physics \cite{6}, and many more. The definition of s-RPA is not uniform, depending on whether exchange is included or not. We understand it, e.g. as in nuclear physics \cite{6}, as the small amplitude limit of time dependent Hartree--Fock theory (TDHF) and therfore with exchange. Its popularity probably stems from its conceptual simplicity, its numerical tractability (in spite of some serious problems in finite size systems), and most of all its well behaved properties conserving full fillment of conservation laws (Ward identies), Goldstone theorem, and restoration of spontaneously broken symmetries. Though there exist respectable general theories (see e.g. \cite{7,8}), any practical attempt to go beyond this basic HF--RPA scheme conserving these properties turned out to be  technically extremely demanding and no well accepted general and practical extension has emerged so far. Nevertheless, standard RPA has also quite serious shortcomings and it is desirable to overcome them. One of the most prominent is its violation of the Pauli principle, often paraphrased as the ``quasiboson approximation". It is most critical for only moderately collective modes or when the self interaction of the gas of quantum fluctuations becomes important as in ultra small finite quantum systems. Since a couple of years two of the present authors and collaborators have been working on a non linear extension of RPA \cite{9} which has shown surprisingly accurate results in a number of non-trivial models \cite{10}. It is called Self --Consistent --RPA (SCRPA) and can be obtained from minimising an energy weighted sum rule. Therefore s- RPA which is perturbative in the sense that it sums a certain class of diagrams (the bubbles) is upgraded in SCRPA to a nonperturbative variational theory though it is in general not of the Raleigh--Ritz type. A strong bonus of this extension of s-RPA is that it generally preserves its positive features as conservation laws and restoration of symmetries as well as numerical tractability, since it leads to equations of the Schr\"odinger type \cite{11}. In this paper we want to apply this theory to the Hubbard model for strongly correlated electrons. Because of its necessarily increased numerical complexity over s-RPA, we first want to consider finite clusters in reduced dimensions. Before going into the details, let us very briefly repeat the main ideas of SCRPA.

One way of presentation is to outline its strong analogy with the Hartree-Fock-Bgoliubov (HFB) approach to interacting boson fields $~b^{\dagger}, ~b$. The HFB canonical transformation reads
\ba
q^{+}_{\nu} =\sum\limits_{i} u_{i, \nu}\; b^{\dagger}_{i} - v_{i, \nu}\; b_{i} \label{qalphaboson}
\ea
The amplitudes $u, ~v$ can be determined \cite{12} from minimising the following mean energy (energy weighted sum rule)
\be
\omega_{\nu} = \frac{\left\langle 0|\left[ q_{\nu} \,,\left[ H\,,q^{\dagger}_{\nu} \right] \right] |0 \right \rangle } {\left\langle 0|\left[ q_{\nu} \,\,,q^{\dagger}_{\nu} \right] |0 \right \rangle } \label{omega-nu_boson}
\ee
where $H$ is the usual many body Hamiltonian with two body interactions and the groundstate $|0 \rangle $ is supposed to be the vacuum to the quasiboson operators $q_{\nu}$, i.e.
\be
q_{\nu} |0 \rangle = 0~.\label{cond_vac_boson}
\ee
With this scheme and the usual orthonormalisation conditions for the amplitudes $u, ~v$, which allows the inversion of (\ref{qalphaboson}), one derives standard HFB theory \cite{6} with no need to construct $|0 \rangle $ explicitly. Of course, in this way the fact that HFB is a Raleigh--Ritz variational theory is not manifest but the scheme has the advantage to be physically transparent and to lead to the final equations with a minimum of mathematical effort.

For SCRPA we follow exactly the same route. We replace in (\ref{qalphaboson}) the ideal boson operators by fermion pair operators of the particle--hole ($ph$) type and form an ansatz for a general transformation of $ph$ -Fermion pairs
\be
Q^{\dagger}_{\nu}= \sum\limits_{ph} \left(\X_{ph}^{\nu}\;a^{\dagger}_{p} a_{h} - \Y_{ph}^{\nu} \; a^{\dagger}_{h} a_{p} \right)~,
\label{ph-operator}
\ee
with $~|\nu\rangle = Q^{\dagger}_{\nu} |0\rangle ~$ an excited state of the spectrum. In analogy with (\ref{omega-nu_boson}) we minimise a mean excitation energy 
\be
\Omega_\nu = \frac {\left\langle 0|\left[ Q_\nu ,\,\left[ H,\,Q_{\nu}^{\dagger}\right]\right]|0 \right \rangle }{\left\langle 0|\left[ Q_\nu ,\,Q_{\nu}^{\dagger}\right]|0 \right\rangle }
\label{energie_nu}
\ee
with $|0 \rangle $, in analogy with (\ref{cond_vac_boson}), the vacuum to the operators $Q_{\nu}$, i.e. 
\be
Q_{\nu} |0 \rangle = 0\label{cond_minimisation-rpa}
\ee
and arrive at equations of the usual RPA type \cite{6}
\ba
\left(  \barr{cc} 
\A &  \B  \cr -\B^{*} & -\A^{*} \earr \right)
\left(  \barr{cc}  \X^{\nu} \cr  \Y^{\nu} \earr \right) = 
\Omega_\nu   \left(  \barr{cc}  \X^{\nu} \cr  \Y^{\nu} \earr \right)
\label{eqSCRPA_matricielle}
\ea
with
\ba
\A_{ph, p'h'} &=& \frac{ \left\langle 0|\left[ a^{\dagger}_{h} \,a_{p} \,,\left[ H,\, a^{\dagger}_{p'} \,a_{h'}\right]\right]|0 \right \rangle }{\sqrt{n_h -n_p}\sqrt{n_{h'} -n_{p'}}} 
\nonumber \\
\nonumber \\
\B_{ph, p'h'} &=& - \frac{ \left\langle 0|\left[ a^{\dagger}_{h} \,a_{p} \,,\left[ H,\, a^{\dagger}_{h'} \,a_{p'}\right]\right]|0 \right \rangle }{\sqrt{n_h -n_p}\sqrt{n_{h'} -n_{p'}}}~. \label{elements-of-RPA}
\ea
Here we supposed to work in a single particle basis which diagonalises the density matrix (natural orbits) :
\be
\langle 0|a^{\dagger}_{k} \,a_{k'}|0 \rangle \equiv n_k \delta_{kk'}\label{number-occup}
\ee
and therefore the $ n_k $'s are the occupation numbers. For $H$ with a two body interaction, (\ref{elements-of-RPA}) only contains correlation functions of the $\langle a^{\dagger}\,a \rangle $ and $\langle a^{\dagger}\,a \,a^{\dagger}\,a \rangle $ types and, since (\ref{cond_minimisation-rpa}) admits the usual RPA orthonormalisation relations for the amplitudes $\X, ~\Y$ \cite{6}, the relation (\ref{ph-operator}) can be inverted and with (\ref{cond_minimisation-rpa}) the correlation functions in (\ref{elements-of-RPA}) be expressed by $\X, ~\Y$.

However, to be complete, occupation numbers $n_k = \langle 0|a^{\dagger}_{k} \,a_{k}|0 \rangle $ and two body correlation functions with other index combinations than two times particle and two times hole need extra considerations. That will be done in the main text. This is, in short, the SCRPA scheme which, as HFB theory, is obviously non linear, since the elements $\A$ and $\B$ in (\ref{eqSCRPA_matricielle}) become functionals of the $\X$ and $\Y$ amplitudes. We want to point out that no bosonisation of Fermion pairs is operated at any stage of the theory.

We want to apply this scheme to the Hubbard model of strongly correlated electrons which is one of the most wide spread models to investigate strong electron correlations and high $T_c$ superconductivity. Its Hamiltonian is given by
\be
  H \,= -t\, \sum \limits_{<i j> \sigma}\,c^{\dagger}_{i \sigma} c_{j \sigma}
  \,+\, U \sum \limits_{i} \hat{n}_{i \uparrow} \hat{n}_{i \downarrow}~\label{HubbardCoordSpace}
\ee
where $~c^{\dagger}_{i \sigma}$, $~c_{j \sigma}~$ are the electron creation and destruction operators at site `$i$' and the $\hat{n}_{i\sigma}=c^{+}_{i\sigma}\,c_{i\sigma}$ are the number operators for electrons at site `$i$' with spin projection $\sigma$. As usual $t$ is the nearest neighbour hopping integral and $U$ the on site coulomb matrix element. In this exploratory work, we will limit ourselves to the simplest cases possible, i.e. we will consider closed chains in one dimension with increasing number of sites at half filling, starting with the 2 -sites problem. It will turn out that the next case of 4 -sites is a configuration with degeneracies which cause problems in SCRPA, as do all $4n$ ($n=1,2,3, \ldots$) configurations in $1D$. We therfore will postpone the treatment of these cases to future work and directly jump to the case of 6 -sites and only shortly outline at the end why the 4 -sites case is unfavorable and how the problem can eventually be cured. In this work we will stop with the 6 -sites case considering it as sufficiently general to be able to extrapolate to the more electron case. In this way one may hope to approach the thermodynamic limit in increasing the number of sites as much as possible. Let us mention that an earlier attempt to solve SCRPA in $1D$ in the thermodynamic limit in a strongly simplified version of SCRPA, the so-called renormalised RPA (r-RPA), produced interesting results \cite{13}.

In detail our paper is organised as follows : in section (\ref{2siteprob}) we present the two sites case with its exact solution. In section (\ref{6siteprob}) we outline the 6 -sites case with a detailed discussion of the results and in section (\ref{4siteprob}) we present the difficulties encountered in the 4 -sites case and how, eventually, one can overcome them. Finally in section (\ref{Conclusions}) we give our conclusions together with some perspectives of this work.

\section{\label{2siteprob} The two sites problem}

In this section we will apply the general formalism of SCRPA outlined in the introduction to the two sites problem at half filling, i.e. two electrons with periodic boundary conditions. This case may seem trivial, the fact, however, is that such popular many body approximations as standard RPA (s-RPA), GW \cite{14}, Gutzwiller wave function \cite{15}, the TPSC approch by Vilk and Tremblay \cite{16}, etc.. do not yield very convincing results in this study case, whereas it has recently been shown that SCRPA solves two body problems exactly \cite{10,11,24}. We again will briefly demonstrate this here for the two sites problem.

First we will transform (\ref{HubbardCoordSpace}) into momentum space. With the usual transformation to plane waves $~c_{j,\sigma}=\frac 1{\sqrt{N}} \sum\limits_{\bf{\vec{k}}} a_{\bf{\vec{k}},\sigma} e^{-i\bf{\vec{k}\,\vec{x}_{j}}}$ this leads to the standard expression for a zero range two body interaction :
\ba
H = & & \sum_{{\bf{\vec{k}}}, \sigma} \left (\epsilon_{k} - \mu \right ) \hat{n}_{{\bf{\vec{k}}}, \sigma}
\nonumber \\
& + & \frac U{2\,N}\sum_{\bf{\vec{k},\vec{p},\vec{q}}, \sigma} a_{{\bf{\vec{k}}}, \sigma}^{\dagger} \, a_{{\bf{\vec{k}+\vec{q}}}, \sigma} \,a_{{\bf{\vec{p}}}, -\sigma}^{\dagger} \, a_{{\bf{\vec{p}-\vec{q}}}, -\sigma}
\label{hamiltonian_imp}
\ea
where $\hat{n}_{\bf{\vec{k}}, \sigma} = a^{\dagger}_{\bf{\vec{k}}, \sigma} a_{\bf{\vec{k}}, \sigma}$ is the occupation number operator of the mode ($\bf{\vec{k}}, \sigma$) and the single particle energies are given by $\epsilon _{\bf{\vec{k}}}= -2\,t\sum\limits_{d=1}^{D } cos\left(k_{d}\right)$ with the lattice spacing set to unity.

For our further considerations it is convenient to transform (\ref{hamiltonian_imp}) to HF quasiparticles operators via (we switch to $1D$)
\be
a_{h, \sigma} = b^{\dagger}_{h,\sigma} ~\mbox{,}~~~~~~~~~~~~~~~~ a_{p, \sigma} = b_{p, \sigma}~\mbox{,}
\ee
where $h$ and $p$ are momenta below and above the Fermi momentum, respectively, so that ${b_{k, \sigma}|HF\rangle }=0$ for all $k$ where ${|HF\rangle }$ is the Hartree--Fock groundstate in the plane wave basis. For the two sites problem with periodic boundary conditions we then write, after normal ordering, the Hamiltonian (\ref{hamiltonian_imp}) in the following way
\be
H = H_{HF} + H_{q=0} + H_{q=\pi}\label{hamilton_2sit}
\ee
with
\ba
H_{HF}&=& E_{HF} + \sum\limits_{\sigma} \left [-\epsilon_{1} \; \tilde{n}_{k_1,\sigma} + \epsilon_{2} \;\tilde{n}_{k_2,\sigma} \right ]
\nonumber \\
\epsilon_{1}&=& -t +\frac{U}{2}~~~~~~~~~~~~~~~~\epsilon_{2}= t +\frac{U}{2}
\label{h1} \\
H_{q=0} &=& \frac{U}{2} \left (\tilde{n}_{k_2,\uparrow} - \tilde{n}_{k_1,\uparrow}\right ) \left (\tilde{n}_{k_2,\downarrow} - \tilde{n}_{k_1,\downarrow}\right )
\label{h2} \\
H_{q=\pi}&=& -\frac{U}{2}\left( J_{\uparrow}^{-} + J_{\uparrow}^{+} )( J_{\downarrow}^{-} + J_{\downarrow}^{+} \right)
\label{h3}
\ea
and $~J^{-}_{\sigma} = b_{1,\sigma}\,b_{2,\sigma}$, $~ J^{+}_{\sigma} = \left(J^{-}_{\sigma}\right)^{+}$, $~\tilde{n}_{k_i,\sigma} = b^{\dagger}_{i,\sigma } \,b_{i,\sigma}~$ where we introduced the abreviation ``1" and ``2" for the two momenta $k_1 =0$ and $k_2 =-\pi $ of the system, respetively. The HF groundstate is $|HF\rangle = b_{1,\uparrow } \,b_{1,\downarrow}|vac \rangle $ and the corresponding energy is given by 
\be
E^{HF}_{0} =\langle {HF}| H |HF\rangle = -2\,t+\frac{U}{2}
\ee
The RPA excitation operator corresponding to (\ref{ph-operator}) can, because of rotational invariance in spin space, be separeted according to spin singlet ($S=0$, charge) and spin triplet ($S=1$) excitations. The latter still can be divided into spin longitudinal ($S=1$, $m_s =0$) and spin transverse ($S=1$, $m_s =\pm 1$) excitations. Let us first consider the charge and spin longitudinal sectors. For later convenience we will not separate them and write for the corresponding RPA operator 
\be
Q^{\dagger}_{\nu} = \X^{\nu}_{\uparrow}\, K^{+}_{\uparrow} + \X^{\nu}_{\downarrow}\,K^{+}_{\downarrow}
- \Y^{\nu}_{\uparrow}\, K^{-}_{\uparrow} - \Y^{\nu}_{\downarrow}\,K^{-}_{\downarrow}
\label{opRPA_excit2sit}
\ee
where $~K^{\pm}_{\sigma}=J^{\pm}_{\sigma}/\sqrt{1 - \langle M_\sigma\rangle}$, $~M_\sigma =\tilde{n}_{1\sigma} + \tilde{n}_{2\sigma} ~$, and the mean values $\langle \ldots \rangle $ are always taken with respect to the RPA vacuum
\be
Q_{\nu} |RPA \rangle = 0 ~.\label{vacuum-rpa}
\ee
Because of the orthonormality relations
\ba
&&\sum\limits_{\sigma}\;\left ( \X^{\nu}_{\sigma} \X^{\nu'}_{\sigma} - \Y^{\nu}_{\sigma} \Y^{\nu'}_{\sigma}\right ) = \delta _{\nu \nu '}~,
\nonumber \\
&&\sum\limits_{\sigma}\;\left ( \X^{\nu}_{\sigma}\Y^{\nu '}_{\sigma} -\Y^{\nu}_{\sigma}\,\X^{\nu '}_{\sigma} \right ) = 0~,
\nonumber \\
&&\sum\limits_{\nu}\;\left ( \X^{\nu}_{\sigma} \X^{\nu}_{\sigma'} - \Y^{\nu}_{\sigma} \Y^{\nu}_{\sigma'}\right ) = \delta _{\sigma \sigma '}~,
\nonumber \\
&&\sum\limits_{\nu}\;\left ( \X^{\nu}_{\sigma}\Y^{\nu}_{\sigma'} -\Y^{\nu}_{\sigma}\,\X^{\nu}_{\sigma'} \right ) = 0~,\label{rel_norm-ferm2sits}
\ea
one can invert (\ref{opRPA_excit2sit}) to obtain
\ba
J^{-}_{\sigma} = \sqrt{1-\langle M_{\sigma}\rangle}\;\sum\limits_{\nu}\;
\left(\;\X^{\nu}_{\sigma} \; Q_{\nu} + \Y^{\nu}_{\sigma} \; Q_{\nu}^{\dagger}\;\right)~,
\nonumber \\
J^{+}_{\sigma} = \left(J^{-}_{\sigma}\right)^{\dagger}~.
\label{inversion_1}
\ea
The operators $J^{\pm}_{\sigma}$ and $1-M_\sigma$ form a $SU(2)$ algebra of spin --$\frac 1 2$ operators and, therfore, using the casimir relation we obtain
\be
M_\sigma = 2\, J^{+}_{\sigma}\, J^{-}_{\sigma}
\ee
In this way we can calculate with (\ref{vacuum-rpa}) the following expectation values
\ba
\left\langle J^{+}_{\sigma'}\,J^{-}_{\sigma}\right\rangle & = & \sqrt{\langle 1- M_{\sigma'}\rangle \langle 1 - M_{\sigma}\rangle}\; \sum\limits_{\nu}\;\Y^{\nu}_{\sigma'}\; \Y^{\nu}_{\sigma}\mbox{,}
\nonumber \\
\left\langle J^{-}_{\sigma'}\,J^{+}_{\sigma}\,\right\rangle & = &\sqrt{\langle 1- M_{\sigma'}\rangle \langle 1 - M_{\sigma}\rangle}\; \sum\limits_{\nu}\;\X^{\nu}_{\sigma'}\; \X^{\nu}_{\sigma}\mbox{,}
\nonumber \\
\left\langle J^{+}_{\sigma'}\,J^{+}_{\sigma}\,\right\rangle & = &\sqrt{\langle 1- M_{\sigma'}\rangle \langle 1 - M_{\sigma}\rangle}\; \sum\limits_{\nu}\;\Y^{\nu}_{\sigma'}\; \X^{\nu}_{\sigma}\mbox{,}
\nonumber \\
\left\langle J^{-}_{\sigma'}\,J^{-}_{\sigma}\,\right\rangle & = &\sqrt{\langle 1- M_{\sigma'}\rangle \langle 1 - M_{\sigma}\rangle}\; \sum\limits_{\nu}\;\X^{\nu}_{\sigma'}\; \Y^{\nu}_{\sigma}\mbox{,}
\label{vm-jj-2sit}
\ea
with
\be
\left\langle M_{\sigma} \right\rangle  = \frac {2\;\sum\limits_{\nu}\;|\Y^{\nu}_{\sigma }|^{2}}{1+ 2\; \sum\limits_{\nu} \;|\Y^{\nu}_{\sigma }|^{2}}\mbox{.}\label{vm-msigma}
\ee
We will see that in order to close the system of SCRPA equations, expectation values $\langle M_{\sigma} M_{\sigma'} \rangle$ will also be needed. It is easy to see that we have 
\be
M_\sigma \,M_\sigma = 2 \,M_\sigma ~
\ee
and
\be
M_{\sigma} M_{\sigma'} = 4 \, J^\dagger_{\sigma} J^\dagger_{\sigma'} J_{\sigma'} J_{\sigma} ~~~~~~(\sigma \neq \sigma') ~. \label{msms}
\ee
With (\ref{inversion_1}) the expectation value of (\ref{msms}) gives
\ba
\langle M_{\sigma} M_{\sigma'} \rangle = 4 (1- \langle M_{\sigma}\rangle) (1- \langle M_{\sigma'} \rangle) 
\nonumber \\
.\sum\limits_{\nu \nu'}\sum\limits_{\nu_1 \nu_2} \Y^{\nu}_{\sigma} \Y^{\nu'}_{\sigma} \Y^{\nu_1}_{\sigma'} \Y^{\nu_2}_{\sigma'} \langle Q_\nu Q_{\nu_1} Q^\dagger_{\nu_2} Q^\dagger_{\nu'}
\rangle ~.\label{mm-2sit}
\ea
For the calculation of the correlation functions which appear on the right hand side of (\ref{mm-2sit}) one commutes the destructors $Q_\nu$ to the right and uses (\ref{cond_minimisation-rpa}), yielding again correlation functions $\langle M_{\sigma} M_{\sigma'} \rangle$. One then obtains a closed linear system of equations for the latters. Details are given in Appendix (\ref{correl-ph}).

The SCRPA matrix elements can be expressed in the following way
\ba
\A_{\uparrow ,\uparrow} = \left\langle \left[ K^{-}_{\uparrow} ,\left[ H,K^{+}_{\uparrow}\right] \right] \right\rangle
&=& 2\,t + \B_{\uparrow ,\uparrow}
\nonumber \\
\A_{\downarrow ,\downarrow}=\left\langle \left[ K^{-}_{\downarrow} ,\left[ H,K^{+}_{\downarrow}\right] \right] \right\rangle
&=& 2\,t + \B_{\downarrow ,\downarrow} 
\nonumber \\
\A_{\uparrow ,\downarrow} = \left\langle \left[ K^{-}_{\uparrow} ,\left[ H,K^{+}_{\downarrow}\right] \right] \right\rangle
&=& \B_{\uparrow ,\downarrow} 
\nonumber \\
\A_{\downarrow ,\uparrow}  = \left\langle \left[ K^{-}_{\downarrow} ,\left[ H,K^{+}_{\uparrow}\right] \right] \right\rangle
&=& \B_{\downarrow ,\uparrow}
\label{elementsA_rpa_bass_sph}
\ea
\ba
\B_{\uparrow ,\uparrow} &=& -\left\langle \left[ K^{-}_{\uparrow} ,\left[ H,K^{-}_{\uparrow}\right] \right] \right\rangle
\nonumber \\
&=& U\, \sqrt{ \frac {1 -\langle M_{\downarrow }\rangle}{1 -\langle M_{\uparrow }\rangle}}\;
\sum\limits_{\nu}\;\left ( \X^{\nu}_{\uparrow} \,\Y^{\nu}_{\downarrow}+\X^{\nu}_{\uparrow} \,\X^{\nu}_{\downarrow}\right )
\nonumber \\
\B_{\downarrow ,\downarrow} &=& -\left\langle \left[ K^{-}_{\downarrow} ,\left[H,K^{-}_{\downarrow}\right] \right] \right\rangle
\nonumber \\
&=&  U\,\sqrt{ \frac {1 -\langle M_{\uparrow }\rangle}{1 -\langle M_{\downarrow }\rangle}}\;
\sum\limits_{\nu}\;\left ( \X^{\nu}_{\uparrow}\,\Y^{\nu}_{\downarrow} +\Y^{\nu}_{\uparrow}\,\Y^{\nu}_{\downarrow} \right )
\nonumber \\
\B_{\uparrow ,\downarrow} &=& -\left\langle \left[ K^{-}_{\uparrow} ,\left[ H,K^{-}_{\downarrow}\right] \right] \right\rangle
\nonumber \\
&=& - \frac U{2} \frac {\langle ( 1 - M_{\uparrow } )(1 - M_{\downarrow })\rangle }{\sqrt {(1 -\langle M_{\uparrow }\rangle )(1 -\langle M_{\downarrow }\rangle )}}
\nonumber \\
\B_{\downarrow ,\uparrow} &=& -\left\langle \left[ K^{-}_{\downarrow} ,\left[ H,K^{-}_{\uparrow}\right] \right] \right\rangle
=\B_{\uparrow ,\downarrow}
\label{elementsB_rpa_bass_sph}
\ea
With our previous relations (\ref{vm-jj-2sit}), (\ref{vm-msigma}) and (\ref{mm-2sit}) we can entirely express the elements of
(\ref{elementsA_rpa_bass_sph}) and (\ref{elementsB_rpa_bass_sph}) by the RPA --amplitudes and therefore we have a completely closed system of equation for the amplitudes $\X,~\Y$. With the orthonormality relations (\ref{rel_norm-ferm2sits}) we further more have
\ba
\A_{\uparrow ,\uparrow} &=& \A_{\downarrow ,\downarrow} = A ~\mbox{,}
~~~~~~~~~
\A_{\uparrow ,\downarrow} = \A_{\downarrow ,\uparrow} = A' ~\mbox{,}
\nonumber \\
\B_{\uparrow ,\uparrow} &=& \B_{\downarrow ,\downarrow} = B ~\mbox{,}
~~~~~~~~~
\B_{\uparrow ,\downarrow} = \B_{\downarrow ,\uparrow} = B' ~\mbox{.}
\ea
and, therefore, the SCRPA equation can be written in the following form
\be
\left(\barr{cccc}  A & A' & B & B' \\ A' & A & B' & B \\
-B & -B' & -A & -A' \\ -B' & -B & -A' & -A  \earr \right)
\left(  \barr{c}  \X^{\nu}_{\uparrow}  \\ \X^{\nu}_{\downarrow}  \\ \Y^{\nu}_{\uparrow} \\ \Y^{\nu}_{\downarrow} \earr \right)
= {\mathcal{E}}_{\nu}\left(   \barr{c}  \X^{\nu}_{\uparrow}  \\ \X^{\nu}_{\downarrow} \\
\Y^{\nu}_{\uparrow} \\ \Y^{\nu}_{\downarrow}   \earr \right) \mbox{.}
\label{rpa11}
\ee
The system (\ref{rpa11}) has the two positive roots ${\E_{1} = \sqrt{ \left(A-A'\right)^2 - \left(B-B'\right)^2}}~$ and ${\E_{2} = \sqrt{ \left(A+A'\right)^2 - \left(B+B'\right)^2}}$. The SCRPA equation (\ref{rpa11}) can be solved numerically by iteration leading, as expected, to the exact result. This latter fact can also be seen analytically in noticing that by symmetry 
\ba
\X^{1}_{\uparrow}& = & -\X^{1}_{\downarrow} \equiv \X_{sp}~\mbox{,}~~~~~~~~
\Y^{1}_{\uparrow} =  -\Y^{1}_{\downarrow} \equiv \Y_{sp}~\mbox{,}
\nonumber \\
\X^{2}_{\uparrow}& = & \X^{2}_{\downarrow} \equiv \X_{ch}~\mbox{,}~~~~~~~~~~
\Y^{2}_{\uparrow} = \Y^{2}_{\downarrow} \equiv \Y_{ch}~\mbox{.}
\label{amp_rpa11}
\ea
Therefore the $4\times 4$ equation (\ref{rpa11}) decouples into two $2\times 2$ equations corresponding to charge ($ch$) and spin ($sp$). Then we see that the exact groundstate wave function which contains only up to $2p-2h$ excitations
\be
|0\rangle \propto \left( 1 + d \; J^{+}_{\uparrow}\, J^{+}_{\downarrow}\right)|HF\rangle \label{groundst}
\ee
is the exact vacuum to the RPA operators, i.e $~Q_{ch (sp)} |RPA \rangle = 0~$ under the condition that
\be
d=\left(\frac{ \Y }{ \X}\right)_{ch (sp)} \equiv \tan\left(\phi\right)~. \label{d}
\ee
We therfore can express the SCRPA equations by the single parameter $\phi$ and obtain the solution analytically (up to the solution  a nonlinear equation for $\phi$). The solution agrees for all quantities with the exact result. For example the groundstate energy is given by
\be
E^{SCRPA}_{0} = -2\,t cos(2\,\phi) + \frac U{2}\; \left (1- sin(2\,\phi) \right )~.
\label{energ_fond_SCRPAxyd}
\ee
This expression can either be derived directly from $\langle H\rangle $ using (\ref{groundst}) and (\ref{d}) or one uses a generalisation of the standard RPA expression for the groundstate energy \cite{6}:
\be
E^{SCRPA}_{0} = E_{HF} -\frac 12\sum\limits_{\sigma} (1-\left\langle M_{\sigma} \right\rangle) \left [\E_{2} |\Y_{ch}|^{2} + \E_{1} |\Y_{sp}|^{2}\right ]~.
\label{energ_fond_SCRPAxy}
\ee
It is straightforward to verify that expressions (\ref{energ_fond_SCRPAxyd}) and (\ref{energ_fond_SCRPAxy}) are identical.

\begin{figure}[ht]
     \includegraphics[width=8cm,height=7cm]{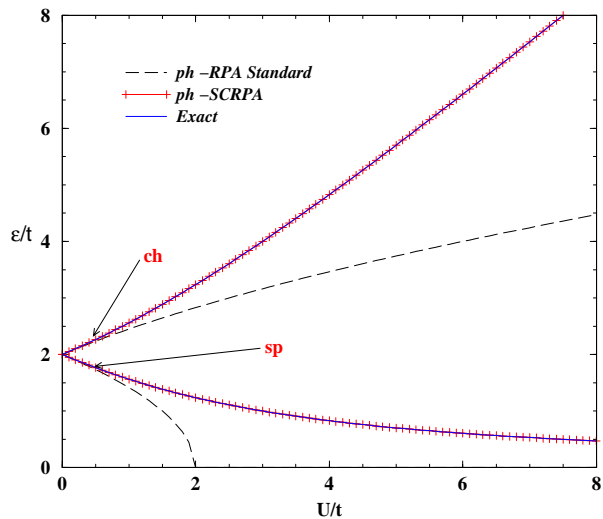}
      \caption{\label{figure1} Excitation energies of the standard RPA (dashed lines), SCRPA (crosses) and exact solution (solid lines) as a function of $U$ in the channels of charge ($ch$) and longitudinal spin ($sp$) for the 2 -sites case.}
\end{figure}

\begin{figure}[ht]
   \includegraphics[width=8cm,height=7cm]{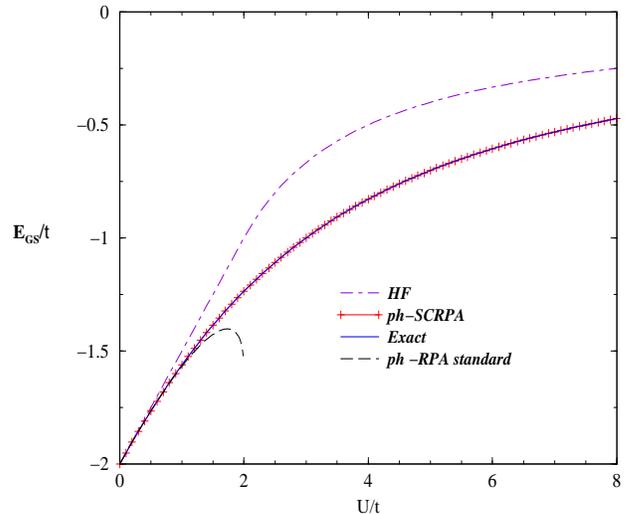}
    \caption{\label{figure2} Groundstate energy in HF (dot-dashed line), standard RPA (dashed line), SCRPA (crosses) and exact solution (solid line) as a function of $U$ in the charge and longitudinal spin responses for the 2 -sites case.}
\end{figure}

The standard RPA expression are recovered from (\ref{rpa11}) in replacing in all expectation values the RPA groundstate by the uncorrelated HF determinant. In Fig.\ref{figure1} we compare the standard RPA with SCRPA and exact results for the excitation energies and in Fig.\ref{figure2} the corresponding groundstate energies together with the HF--values are shown. From these figures one should especially appreciate the long way SCRPA has gone from s-RPA to recover the exact result. For instance it is clearly seen that the instability of s-RPA at $U=2$ is, as expected for such a small system, an artefact and is completely washed out by the self consistent treatment of quantum fluctuation contained in the SCRPA approach.

Without explicit demonstration let us also mention that SCRPA in the spin transverse channel with
$Q^{\dagger}_{\nu} = \X^{\nu}_{1\downarrow 2\uparrow}\, b^{\dagger}_{2\uparrow} \, b^{\dagger}_{1\downarrow} + \X^{\nu}_{1\uparrow 2\downarrow}\, b^{\dagger}_{2\downarrow} \, b^{\dagger}_{1\uparrow} -\Y^{\nu}_{1\downarrow 2\uparrow}\, b_{1\downarrow}\, b_{2\uparrow} - \Y^{\nu}_{1\uparrow 2\downarrow}\, b_{1\uparrow}\, b_{2\downarrow} $ as well as in the particle--particle channel with $Q^{\dagger} = \X\, b^{\dagger}_{2\uparrow} \, b^{\dagger}_{2\downarrow} - \Y\, b_{1\downarrow}\, b_{1\uparrow} $ also gives the exact solution for the two sites problem. How the $pp$--SCRPA works can be seen in ref\cite{10} where for the pairing problem the two particle problem is also solved exactly.

The fact that SCRPA solves the two sites problem exactly is non trivial, since other well known many body approaches \cite{14,15,16}, as already mentioned, so far failed to obtain this limit correctly.

\section{\label{6siteprob} The six sites problem}

\begin{figure}[ht]
  \includegraphics[width=6.5cm,height=4.5cm]{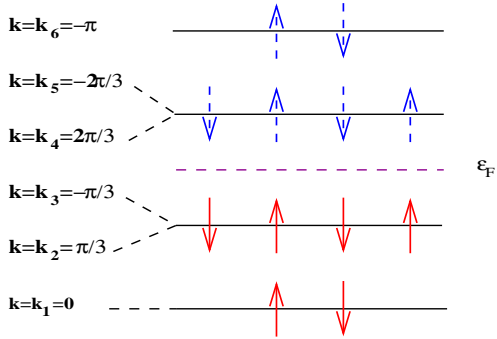}
      \caption{\label{figure3} Excitation spectrum of HF at $U=0$ for the chain with 6 -sites at half filling and projection of spin $m_{s} = 0$. The occupied states are represented by the full arrows and those not occupied are represented by the dashed arrows.}
\end{figure}

After this positive experience with the two sites problem we next will consider the $1$-dimensional 6 -sites case, as for the 4 -sites case problems appear needing particular considerations to be outlined in section \ref{4siteprob}. We again consider the plane wave transformation explained in section \ref{2siteprob} with the corresponding Hamiltonian in momentum space (\ref{hamiltonian_imp}). In the first Brillouin zone $-\pi\leq k < \pi$ we have for $N=6$ the following wave numbers 
\ba
k_{1}&=&0~,~~~~k_{2}=\frac{\pi}{3}~,~~~~k_{3}=-\frac{\pi}{3}~,
\nonumber \\
k_{4}&=&\frac{2\pi}{3}~,~~~~k_{5}=-\frac{2\pi}{3}~,~~~~k_{6}=-\pi~.
\ea
With the HF transformation 
\be
a_{h, \sigma} = b^{\dagger}_{h,\sigma} \mbox{,}~~~~~~~~ a_{p, \sigma} = b_{p, \sigma} ~\mbox{,}
\ee
such that $b_{k, \sigma}|HF\rangle =0$ for all $k$, we can write the hamiltonian in the following way (normal order with respect to $b^{\dagger}$, $b$)
\be
H= H_{HF} + H_{|q|=0} +  H_{|q|=\frac{\pi}{3}} + H_{|q|=\frac{2\pi}{3}} + H_{|q|=\pi}\label{ham_quasip-6sit}
\ee
where
\begin{widetext}
\sba
\label{ham_quasip6sita}
\ba
&&H_{HF} = E_{0}^{HF} + \sum\limits_{\sigma} \left ( \epsilon_{4}\, \tilde{n}_{4,\sigma} + \epsilon_{5}\, \tilde{n}_{5,\sigma} +\epsilon_{6}\, \tilde{n}_{6,\sigma}- \epsilon_{1}\,\tilde{n}_{1,\sigma} - \epsilon_{2}\, \tilde{n}_{2,\sigma} - \epsilon_{3}\, \tilde{n}_{3,\sigma} \right ) \label{HF}
 \\
&&H_{|q|=0} = G \sum\limits_{i=1}^{3}\left (\tilde{n}_{p_{i},\uparrow} - \tilde{n}_{h_{i},\uparrow} \right )\sum\limits_{j=1}^{3}\left (\tilde{n}_{p_{j},\downarrow} - \tilde{n}_{h_{j},\downarrow} \right ) \label{nn}
\\
&&H_{|q|=\frac{\pi}{3}} = G \biggl\{\biggl\{\left [\left(S^{-}_{4\uparrow,6\uparrow} + S^{+}_{6\uparrow,5\uparrow} \right) -\left(S^{+}_{2\uparrow,1\uparrow} + S^{-}_{1\uparrow,3\uparrow}\right) +\left(J^{-}_{2\uparrow,4\uparrow} + J^{+}_{5\uparrow,3\uparrow}\right)\right ] \label{qp3}
\nonumber  \\
&&~~~~~~~~~~~~~~~.\left [\left(S^{+}_{6\downarrow,4\downarrow} + S^{-}_{5\downarrow,6\downarrow}\right) -\left(S^{-}_{1\downarrow,2\downarrow} + S^{-}_{1\downarrow,3\downarrow}\right) + \left(J^{+}_{4\downarrow,2\downarrow} + J^{-}_{3\downarrow,5\downarrow}\right)\right ]\biggr \} + cc \biggr \}
 \\
&&H_{|q|=\frac{2\pi}{3}} = G \biggl\{ \biggl\{\left [\left(S^{+}_{5\uparrow,4\uparrow} -S^{+}_{3\uparrow,2\uparrow} \right) + \left(J^{-}_{1\uparrow,5\uparrow} + J^{+}_{4\uparrow,1\uparrow} + J^{-}_{3\uparrow,6\uparrow} + J^{+}_{6\uparrow,2\uparrow}\right)\right ]
\nonumber \\
&&~~~~~~~~~~~~~~~.\left [\left(S^{-}_{4\downarrow,5\downarrow} -S^{-}_{2\downarrow,3\downarrow} \right) + \left(J^{+}_{5\downarrow,1\downarrow} + J^{-}_{1\downarrow,4\downarrow}+ J^{+}_{6\downarrow,3\downarrow} + J^{-}_{2\downarrow,6\downarrow}\right)\right ]\biggr \} + cc \biggr \}
\label{q2p3}
 \\
&&H_{|q|=\pi} = G \left [\left(J^{-}_{1\uparrow,6\uparrow} + J^{-}_{2\uparrow,5\uparrow} + J^{-}_{3\uparrow,4\uparrow} \right) + cc \right ] \left [\left( J^{-}_{1\downarrow,6\downarrow} + J^{-}_{2\downarrow,5\downarrow} + J^{-}_{3\downarrow, 4\downarrow} \right)+ cc \right ] \label{qp}
\ea
\sea
\end{widetext}
with the following definition of operators
\ba
\barr{ccc}
\tilde {n}_{k, \sigma} = b^{\dagger}_{k, \sigma} b_{k, \sigma}  & & \\
J^{-}_{ph,\sigma}= b_{h, \sigma} b_{p, \sigma} & & J^{+}_{ph,\sigma}=(J^{-}_{ph,\sigma})^{\dagger}  \\
S^{+}_{ll', \sigma}=b^{\dagger}_{l, \sigma} b_{l', \sigma} & \mbox{with}~~~ l>l'  & S^{-}_{l'l,\sigma}=\left(S^{+}_{ll',\sigma}\right)^{\dagger} \earr
\ea
and
\ba
&&E_{HF} = -8\,t + \frac3{4}U \mbox{,}
\nonumber \\
&&\epsilon_{1} = -2\,t + \frac{U}{2}\mbox{,}~~~~~~~~~~~~~~~~\epsilon_{2}=\epsilon_{3}=-t + \frac{U}{2}~\mbox{,}
\nonumber \\
&&\epsilon_{4} = \epsilon_{5} = t + \frac{U}{2}~\mbox{,}~~~~~~~~~~~~~\epsilon_{6}= 2\,t + \frac{U}{2}\mbox{,}
\nonumber \\
&&G=\frac U{6}\mbox{.}
\ea
The level scheme is shown in Fig.\ref{figure3}. 
The hole states are labeled $h=\{1,2,3\}$ and the particle states $p=\{4,5,6\}$. The HF groundstate is
\be
|HF\rangle = a^{\dagger}_{1,\uparrow} \, a^{\dagger}_{1,\downarrow} \, a^{\dagger}_{2,\uparrow}\, a^{\dagger}_{2,\downarrow} \, a^{\dagger}_{3,\uparrow}\, a^{\dagger}_{3,\downarrow} | - \rangle ~.\label{HF-groundstate}
\ee

We see that the Hamiltonian for 6 -sites has largely the same structure as the one for 2 -sites. It is only augmented by $H_{|q|=\frac{\pi}{3}} + H_{|q|=\frac{2\pi}{3}}$ which contains the $S$-operators on which we will comment below.

There are three different absolute values of momentum transfers as shown in Table \ref{table1}.
\begin{table}
\begin{tabular}{|c|c|c|}
\hline
$|q| =\frac{2\pi}{3}$~~~~~~~  & $|q|  =\pi$~~~~~~~ & $|q| = \frac{\pi}{3}$
\\ \hline
$51\rightarrow q_{51} =-\frac{2\pi}{3}$~~~~~~~  & $61\rightarrow q_{61}  =-\pi$~~~~~~~ & $42\rightarrow q_{42} = + \frac{\pi}{3}$
\\ \hline
$41\rightarrow q_{41} =+\frac{2\pi}{3}$~~~~~~~ & $52\rightarrow q_{52}  =-\pi$~~~~~~~ & $53\rightarrow q_{53} =-\frac{\pi}{3}$
\\ \hline
$62\rightarrow q_{62} =+\frac{2\pi}{3}$~~~~~~~ & $43\rightarrow q_{43}  = +\pi$~~~~~~~ &
\\ \hline
$63\rightarrow q_{63} =-\frac{2\pi}{3}$~~~~~~~ & ~~~~~~~ &
\\ \hline
\end{tabular}
\caption{\label{table1} The various momentum transfers in the 6 -sites case.}
\end{table}
Since the momentum transfer $|q|$ is a good quantum number, the RPA equations are block diagonal and can be written down for each $|q|$ -value separately. For exemple for $|q|=\frac{\pi}{3}$ we have the following RPA operator for charge and longitudinal spin excitations
\ba
Q^{\dagger}_{|q|=\frac{\pi}{3},\nu} = &&\X^{\nu}_{2\uparrow, 4\uparrow}\, K^{+}_{4\uparrow,2\uparrow} +\X^{\nu}_{2\downarrow, 4\downarrow}\,K^{+}_{4\downarrow, 2\downarrow} 
\nonumber \\
&+&\X^{\nu}_{3\uparrow, 5\uparrow}\, K^{+}_{5\uparrow,3\uparrow} +\X^{\nu}_{3\downarrow, 5\downarrow}\,K^{+}_{5\downarrow, 3\downarrow}
\nonumber \\
&-&\Y^{\nu}_{2\uparrow, 4\uparrow}\, K^{-}_{2\uparrow,4\uparrow} - \Y^{\nu}_{2\downarrow, 4\downarrow} \,K^{-}_{2\downarrow, 4\downarrow} 
\nonumber \\
&-&\Y^{\nu}_{3\uparrow, 5\uparrow}\, K^{-}_{5\uparrow,3\uparrow} - \Y^{\nu}_{3\downarrow, 5\downarrow}\,K^{-}_{3\downarrow, 5\downarrow}~
\label{op-dexqp6s4}
\ea
where
\be
K^{\pm}_{p\sigma, h\sigma} = \frac{J^{\pm}_{p\sigma, h\sigma}}{\sqrt{1 - \langle M_{p\sigma, h\sigma}\rangle}}
\ee
and 
\be
M_{p\sigma, h\sigma} =\tilde{n}_{p,\sigma} +\tilde{n}_{h,\sigma}~.
\ee
We write this RPA operator in short hand notation as 
\be
Q^{\dagger}_{\nu}= \sum\limits_{i=1}^{4}\frac {1}{\sqrt{1 -\langle M_{i}\rangle}} \left(\X_{i}^{\nu}\;J^{+}_{i} - \Y_{i}^{\nu}\; J^{-}_{i} \right)
\label{op-ex-6sit}
\ee
again with the properties
\sba
\label{prop-vacca}
\ba
|\nu \rangle = Q^{\dagger}_{\nu} |0\rangle \label{nu}
 \\
Q_{\nu} |0\rangle = 0 ~.\label{zero}
\ea
\sea
The matrix elements in the SCRPA equation $\left(\barr{cc} \A &  \B  \\ -\B^{*} & -\A^{*} \earr\right)
\left(\barr{c} \X^{\nu} \\  \Y^{\nu} \earr\right) = \E_\nu   \left(\barr{c} \X^{\nu} \\  \Y^{\nu} \earr\right)$
are then of the form
\sba
\label{eleltsmatix-RPAa}
\ba
\A_{i , i'} &=& \frac {\left\langle \left[ J^{-}_{i'} \left[ H, J^{+}_{i}\right] \right]\right\rangle} {\sqrt{(1-\langle M_{i'}\rangle)( 1-\langle M_{i}\rangle )}}~\mbox{,}\label{A}
 \\
\B_{i, i'} &=& -\frac {\left\langle \left[ J^{-}_{i'} \left[ H, J^{-}_{i}\right] \right] \right\rangle} {\sqrt{(1-\langle M_{i'}\rangle)( 1-\langle M_{i}\rangle )}}~.\label{B}
\ea
\sea

Since the SCRPA equations have the same mathematical structure as standard RPA, one also has equivalent orthonormality relations 
$\sum\limits_{i}\;\left ( \X^{\nu}_{i}\,\X^{\nu'}_{i} - \Y^{\nu}_{i}\, \Y^{\nu'}_{i}\right )
= \delta _{\nu \nu '}~$, etc.. , in analogy to eqs. (\ref{rel_norm-ferm2sits}) of the 2 -sites case. 
This allows to invert (\ref{op-ex-6sit}) and to calculate the expectation values which will appear in (\ref{A}) and (\ref{B}) in complete analogy to (\ref{vm-jj-2sit}).

The missing expectation values $\langle M_i\rangle $ can be expressed by the $\X,\Y$ amplitudes in observing that $J^{\pm}_{i}$, and $J^{0}_{i} = \frac{1}{2} (M_{i} - 1 ) $ form, as in the 2 -sites case, an $SU2$ Lie algebra for spin-$\frac{1}{2}$ particles. Using the Casimir relation one again obtains $M_{i} = 2 \,J^{+}_{i}\,J^{-}_{i} $ and thus
\be
\langle M_{i}\rangle = \frac {2\;\sum\limits_{\nu} |\Y_{i}^{\nu}|^{2}}{1+ 2\;\sum\limits_{\nu} |\Y_{i}^{\nu}|^{2}}~. \label{val-exp-Miph}
\ee
We also will need expectation values of 
\ba
M_i M_{j} =  4 \, J^{+}_i J^{-}_{j} J^{+}_{j}  J^{-}_i  ~~~~~~~\hbox{for~~} i\neq j ~,\nonumber
\ea
(for $M_i M_{i} = 2\,M_{i} $ we can use (\ref{val-exp-Miph})). Those can again be calculated following the same procedure as outlined in (\ref{mm-2sit}) and Appendix (\ref{correl-ph}).

In order to solve the SCRPA equations we now practically have prepared all we need. Nontheless, at this point we have to discuss a limitation of our RPA ansatz (\ref{op-dexqp6s4}) which is not absolutely necessary but which turned out to be convenient for numerical reasons. The fact is that our RPA ansatz is restricted to $ph$ and $hp$ configurations, as this is also the case in standard RPA. In the latter case this is a strict consequence of the use of HF occupation numbers $n^{0}_{p}$, $n^{0}_{h}$ with values zero or one, respectively. In the SCRPA case with a correlated groundstate the occupation numbers are different from zero and one and a priori there is no formal reason not to include into the RPA operator also $pp$ and $hh$ configuration of the form  $a^{\dagger}_{p} a_{p'}\equiv b^{\dagger}_{p} b_{p'} $ and $a^{\dagger}_{h} a_{h'}\equiv - b^{\dagger}_{h'} b_{h} $. Such terms are usually called scattering or anomalous terms \cite{18}. With rounded occupation numbers the SCRPA equations (at $T=0$) are formally and mathematically equivalent to standard RPA equations at finite temperature where also $pp$ and $hh$ components are to be included, in principle \cite{17}. The inclusion of those scattering terms \cite{17,18} (the $S$-terms in (\ref{ham_quasip-6sit})) usually is of little quantitative consequence \cite{11}, entails, however, the important formal property that, as for standard RPA, the energy weighted sum rule is fullfilled exactly \cite{11,18}. Inspite of this desirable feature, we had to refrain from the inclusion of the scattering configurations in this work because the factors $\sqrt{1- \langle M_{i} \rangle }$ by which the SCRPA matrix is divided (see eqs. (\ref{A}) and (\ref{B})), can become very small in these cases and this perturbed the convergence process of the iterative solution of the SCRPA equations. Though we do not exclude that a more adequate numerical procedure could be found to stabilise the iteration cycle, we decided to postpone such an investigation, because, as already mentioned and as will be shown later, the influence of the scattering terms is, as found already in other studies \cite{11}, very small. We will shortly come back to this discussion when presenting the results for the energy  weighted sum rule below. As a consequence and for consistency we then also will have to disregard the $S$-terms of the Hamiltonian (remember that also in standard RPA these terms do not contribute). Under these conditions we then obtain a completely closed system of SCRPA equations. For completeness we give some examples of SCRPA matrix elements which correspond to the ansatz (\ref{op-dexqp6s4}) for $|q|=\frac{\pi}{3}$
\begin{widetext}
\sba
\label{qques-elets6sitsa}
\ba
\A_{1,1} &=& \frac {\left\langle \left[ J^{-}_{2\uparrow,4\uparrow} \left[ H, J^{+}_{4\uparrow,2\uparrow}\right] \right]\right\rangle} {\left(1-\langle M_{24,\uparrow}\rangle\right )}
\nonumber \\
\nonumber \\
&=& \epsilon_{4} -\epsilon_{2} - G \biggl\{ 2\,\langle J^{-}_{2\uparrow,4\uparrow} \left( J^{-}_{3\downarrow,5\downarrow} + J^{+}_{4\downarrow,2\downarrow} \right)\rangle
+ \langle \left (J^{-}_{1\uparrow,4\uparrow} + J^{-}_{2\uparrow,6\uparrow} \right )\left( J^{-}_{1\downarrow,5\downarrow} + J^{-}_{3\downarrow,6\downarrow} + J^{+}_{4\downarrow, 1\downarrow} + J^{+}_{6\downarrow, 2\downarrow} \right) \rangle
\nonumber \\
&&~~~~~~~~~~~~~~~~+\langle\left(J^{-}_{3\uparrow,4\uparrow} + J^{-}_{2\uparrow,5\uparrow} \right) \left [\left( J^{-}_{1\downarrow, 6\downarrow} + J^{-}_{2\downarrow,5\downarrow} + J^{-}_{3\downarrow, 4\downarrow} \right)+ cc \right ]\rangle \biggr\}
.\left(1-\langle M_{24,\uparrow}\rangle\right )^{-1} \label{A11}
\\
\nonumber \\
\A_{2,1}&=&\frac {\left\langle \left[ J^{-}_{2\downarrow,4\downarrow} \left[ H, J^{+}_{4\uparrow,2\uparrow}\right] \right]\right\rangle} {\sqrt{\left(1-\langle M_{24,\downarrow}\rangle\right )\left(1-\langle M_{24,\uparrow}\rangle \right )}}
\nonumber \\
\nonumber \\
&=& G \biggl\{\langle \left ( 1 - M_{24,\uparrow} \right)\left( 1 -  M_{24,\downarrow} \right)\rangle + \langle \left( J^{+}_{4\uparrow,1\uparrow} - J^{+}_{6\uparrow,2\uparrow} \right ) \left( J^{-}_{1\downarrow,4\downarrow} - J^{-}_{2\downarrow,6\downarrow} \right)\rangle
\nonumber \\
&&~~~ + \langle \left( J^{+}_{4\uparrow,3\uparrow} - J^{+}_{5\uparrow,2\uparrow} \right ) \left( J^{-}_{3\downarrow,4\downarrow} - J^{-}_{2\downarrow,5\downarrow} \right)\rangle \biggr\}
\left\{\left(1-\langle M_{24,\downarrow}\rangle\right )\left(1-\langle M_{24,\uparrow}\rangle \right ) \right\}^{-\frac{1}{2}}
\label{A21}
 \\
&& \vdots \nonumber
\ea
\sea
\end{widetext}

The other matrix elements can be elaborated along the same lines. Of course in the approximation where the expectation values in (\ref{A11}) and (\ref{A21}) are evaluated with the HF groundstate the usual matrix elements of standard RPA are recovered. We should also mention that in expressions (\ref{A11}) and (\ref{A21}) expectation values as for example $~\langle J^{-}_{1\uparrow,4\uparrow} \, J^{+}_{4\downarrow, 1\downarrow} \rangle ~$ which involve momentum transfers other than the one under consideration ($|q_3|=\frac{\pi}{3}$ in the specific example) must be discarded. That this implicit channel coupling cannot be taken into account without deteriorating the quality of the SCRPA solutions is an empirical law which has been established quite sometime ago \cite{19}. It is part of the decoupling scheme and it is intuitively understandable that, since each channel is summing specific correlations, one can not mix the channels implicitly without perturbing the balance of the minimisation procedure which is done channel by channel. It can also be noticed that, neglecting the $S$-terms in $H$, the channel coupling disappears.

We here give for the transfer $|q|=\frac{\pi}{3}$ the totality of the elements of matrix SCRPA $\A$ and $\B$ just as it is was used in the numerical calculation. For others transfers there will be analogous expressions. Indeed with the following abbreviations
\ba
i=1\equiv (2\uparrow, 4\uparrow) ~~~~~~~~~ i=2\equiv (2\downarrow, 4\downarrow) 
\nonumber \\
i=3\equiv (3\uparrow, 5\uparrow) ~~~~~~~~~ i=4\equiv (3\downarrow, 5\downarrow)
\nonumber
\ea
the elements of matrices $\A$ and $\B$ are given by
\sba
\ba
\A_{1,1} &=&\epsilon_{4} -\epsilon_{2} - 2\,G  \frac{\langle J^{-}_{2\uparrow,4\uparrow} ( J^{-}_{3\downarrow,5\downarrow} + J^{+}_{4\downarrow,2\downarrow} )\rangle }{ 1-\langle M_{24,\uparrow}\rangle }~,
\nonumber \\
\nonumber \\
\A_{2,1} &=& G \frac {\langle \left ( 1 - M_{24,\uparrow} \right)\left( 1 -  M_{24,\downarrow} \right)\rangle }{\sqrt{\left(1-\langle M_{24,\downarrow}\rangle\right )\left(1-\langle M_{24,\uparrow}\rangle \right ) }}~,
\nonumber \\
\nonumber \\
\A_{3,1} &=& \A_{4,1} = \A_{3,2} = \A_{4,2} =0 ~,
\nonumber \\
\nonumber \\
\A_{2,2} &=& \epsilon_{4} -\epsilon_{2} - 2\,G  \frac{\langle ( J^{-}_{3\uparrow,5\uparrow} + J^{+}_{4\uparrow,2\uparrow} ) J^{-}_{2\downarrow,4\downarrow}\rangle }{ 1-\langle M_{24,\downarrow}\rangle }~,
\nonumber \\
\nonumber \\
\A_{3,3} &=& \epsilon_{5} -\epsilon_{3} - 2\,G  \frac{\langle J^{-}_{3\uparrow,5\uparrow} ( J^{-}_{2\downarrow,4\downarrow} + J^{+}_{5\downarrow,3\downarrow} )\rangle }{ 1-\langle M_{35,\uparrow}\rangle }~,
\nonumber \\
\nonumber \\
\A_{4,3}&=& G \frac {\langle \left ( 1 - M_{35,\uparrow} \right)\left( 1 -  M_{35,\downarrow} \right)\rangle }{\sqrt{\left(1-\langle M_{35,\downarrow}\rangle\right )\left(1-\langle M_{35,\uparrow}\rangle \right ) }}~,
\nonumber \\
\nonumber \\
\A_{4,4} &=& \epsilon_{5} -\epsilon_{3} - 2\,G  \frac{\langle ( J^{-}_{2\uparrow,4\uparrow} + J^{+}_{5\uparrow,3\uparrow} ) J^{-}_{3\downarrow,5\downarrow}\rangle }{ 1-\langle M_{35,\downarrow}\rangle }~,
 \label{eletsA6sits_qps3}
 \\
\nonumber \\
\B_{1,1} &=& - 2\,G  \frac{\langle J^{-}_{2\uparrow,4\uparrow} ( J^{-}_{2\downarrow,4\downarrow} + J^{+}_{5\downarrow,3\downarrow} ) \rangle }{ 1-\langle M_{24,\uparrow}\rangle }~,
\nonumber \\
\nonumber \\
\B_{2,1} &=& \B_{3,1}= \B_{4,2}= \B_{4,3}=0~,
\nonumber \\
\nonumber \\
\B_{4,1}&=& G \frac {\langle \left ( 1 - M_{24,\uparrow} \right)\left( 1 -  M_{35,\downarrow} \right)\rangle }{\sqrt{\left(1-\langle M_{35,\downarrow}\rangle\right )\left(1-\langle M_{24,\uparrow}\rangle \right ) }}~,
\nonumber \\
\nonumber \\
\B_{2,2} &=& - 2\,G  \frac{\langle ( J^{-}_{2\uparrow,4\uparrow} + J^{+}_{5\uparrow,3\uparrow} ) J^{-}_{2\downarrow,4\downarrow}\rangle }{ 1-\langle M_{24,\downarrow}\rangle }~,
\nonumber \\
\nonumber \\
\B_{3,2}&=& G \frac {\langle \left ( 1 - M_{35,\uparrow} \right)\left( 1 -  M_{24,\downarrow} \right)\rangle }{\sqrt{\left(1-\langle M_{24,\downarrow}\rangle\right )\left(1-\langle M_{35,\uparrow}\rangle \right ) }}~,
\nonumber \\
\nonumber \\
\B_{3,3} &=& - 2\,G  \frac{\langle J^{-}_{3\uparrow,5\uparrow} ( J^{-}_{3\downarrow,5\downarrow} + J^{+}_{4\downarrow,2\downarrow} ) \rangle }{ 1-\langle M_{35,\uparrow}\rangle }~,
\nonumber \\
\nonumber \\
\B_{4,4} &=& - 2\,G  \frac{\langle ( J^{-}_{3\uparrow,5\uparrow} + J^{+}_{4\uparrow,2\uparrow} ) J^{-}_{3\downarrow,5\downarrow}\rangle }{ 1-\langle M_{35,\downarrow}\rangle }~.
 \label{eletsB6sits_qps3}
\ea
\sea
Let us add that the matrices $\A$ and $\B$ are symmetric and that the expectation values $\langle \ldots\rangle $ in (\ref{eletsA6sits_qps3}) and (\ref{eletsB6sits_qps3}) can be expressed in an analogous way as the expectation values (\ref{vm-jj-2sit}) and (\ref{mm-2sit}) by the amplitudes $\X, ~ \Y$.

The structure of the self consistent matrix elements (\ref{eletsA6sits_qps3}) and (\ref{eletsB6sits_qps3}) is also quite transparent : the bare interaction which survives in the limit of standard RPA is renormalised, i.e. screened, by two body correlation functions which are calculated  self consistently. The general structure of the scheme is in a way similar to the one proposed by Tremblay and coworkers \cite{16}, however, the details of the expressions and the spirit of derivation are different. One can also interpret our theory as a mean field theory of quantum fluctuations as this was done in refs.\cite{9} 
\begin{figure}[ht]
  \includegraphics[width=8cm,height=7cm]{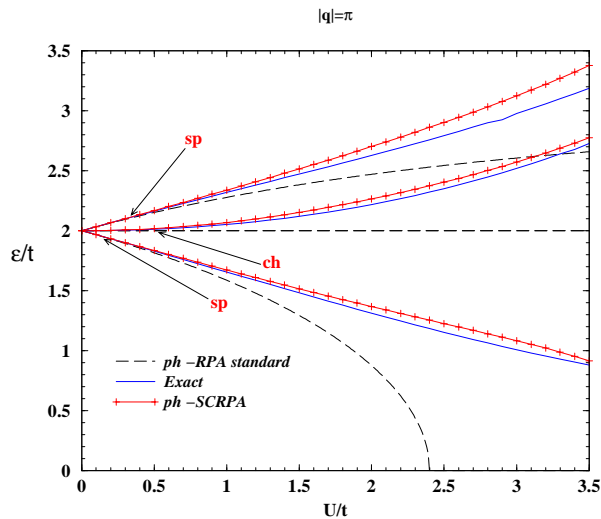}
      \caption{\label{figure4} Energies of excited states in standard RPA, SCRPA, and exact cases as a function of $U$ for 6 -sites with spin projection $m_{s} = 0$ and for $|q|=\pi$. States of the charge response and those of the longitudinal spin response are denoted by $ch$ and $sp$, respectively.}
\end{figure}
\begin{figure}[ht]
  \includegraphics[width=8cm,height=7cm]{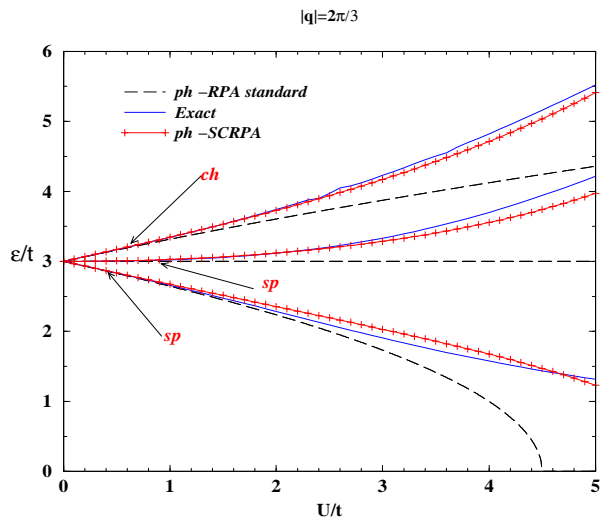}
      \caption{\label{figure5} Same as Fig. \ref{figure4} but for $|q|= \frac {2\pi}{3}$.}
\end{figure}

\begin{figure}[ht]
  \includegraphics[width=8cm,height=7cm]{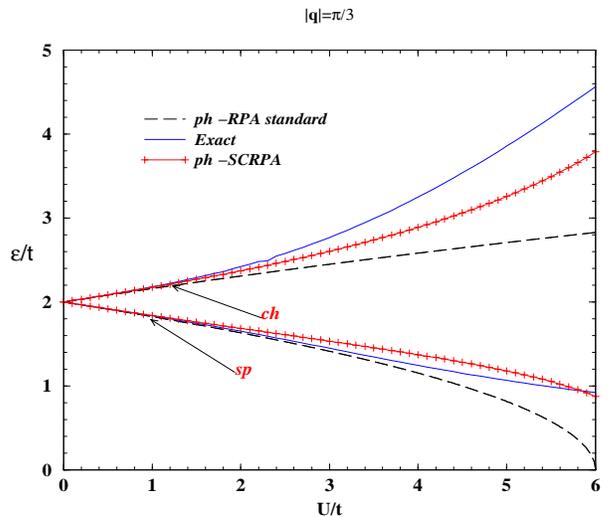}
      \caption{\label{figure6} Same as Fig. \ref{figure4} but for $|q|= \frac {\pi}{3}$}
\end{figure}

Let us now come to the presentation of the results. In Fig. (\ref{figure4}, \ref{figure5} and \ref{figure6}) we display the excitation energies in the three channels $|q|=\pi$, $\frac{2\pi}{3}$ and $\frac{\pi}{3}$ as a fonction of $U/t$. The exact values are given by the continuous lines, the SCRPA ones by crosses and the ones corresponding to standard RPA by the broken lines. We see that in all three cases SCRPA results are excellent and strongly improve over standard RPA. As expected, this is particularly important at the phase transition points where the lowest root of standard RPA goes to zero, indicating the onset of a staggered magnetisation on the mean field level. It is particularly interesting that SCRPA allows to go beyond the mean field instability point. However, contrary to the two sites case where SCRPA, in the plane wave basis, solved the model for all values of $U$, here at some values $U>U_{cr}$ the system ``feels" the phase transition and SCRPA stops to converge and also deteriorates in quality. Up to these values of $U$ SCRPA shows very good agreement with the exact solution and in particular it completely smears the sharp phase transition point of standard RPA which is an artefact of the linearisation.

\begin{figure}[ht]
  \includegraphics[width=8cm,height=7cm]{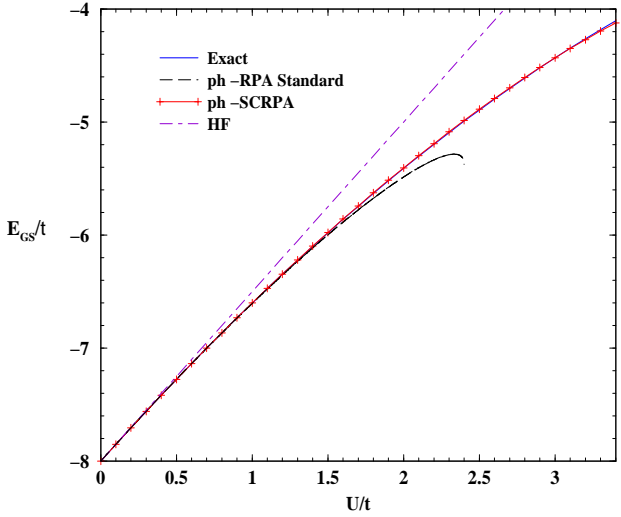}
      \caption{\label{figure7} Energy of groundstate in  HF, standard RPA, SCRPA and exact cases as a function of $U$ for 6 -sites with spin projection $m_{s} = 0$.}
\end{figure}

In Fig. \ref{figure7} we show the groundstate energy (see eq. (\ref{energ_fond_SCRPAxy}))
\be
E^{SCRPA}_{0} =  E_{HF} -\sum\limits_{\nu} \E_{\nu} \sum\limits_{i} (1-\left\langle M_{i} \right\rangle)|\Y^{\nu}_{i} |^{2}
\ee
as a function of $U$. In addition to exact, SCRPA, and s-RPA values we also show the HF energy. Again we see that SCRPA is in excellent agreement with the exact solution. Standard RPA is also good for low values of $U$ but strongly deteriorates close to the lowest phase transition point which occurs in the $|q|=\pi$ channel at $U=\frac{12\,t}{5}$. The HF energies, on the contrary, deviate quite strongly from the exact values. 

\begin{figure}[ht]
  \includegraphics[width=8cm,height=7cm]{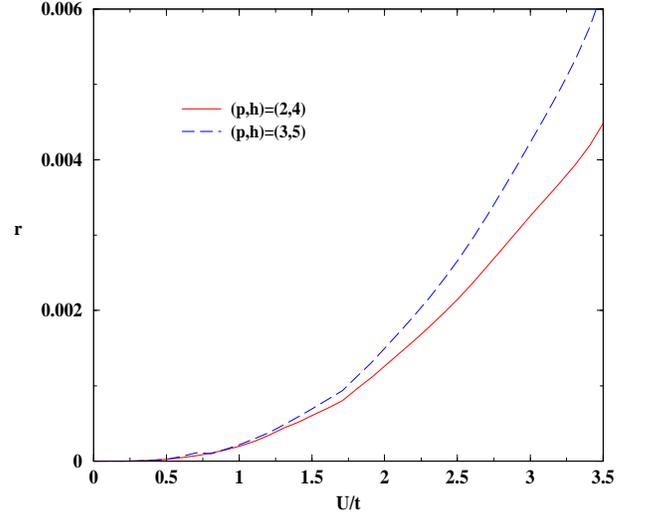}
      \caption{\label{figure8} The ratio, $r$ (eq. \ref{rapp_amplitudes-ud}), as function of the interaction $U$ for the $ph$ excitations $(2,4)$ and $(3,5)$ in the channel $|q|=\frac{\pi}{3}$.}
\end{figure}

\begin{figure}[ht]
  \includegraphics[width=8cm,height=7cm]{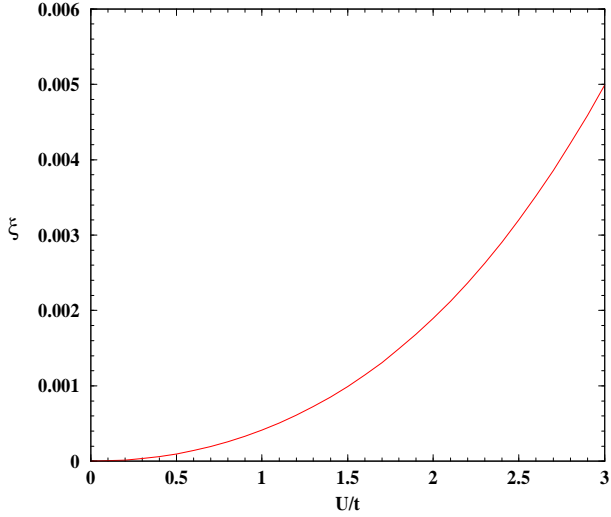}
      \caption{\label{figure9} The ratio, $\xi=\frac{R-L}{R}$, of the energy weighted sum rule in the charge response for the 6-sites case.}
\end{figure}

The reader certainly has remarked that our RPA ansatz (\ref{op-dexqp6s4}) has so far not separated charge and spin excitations. In the 2 -sites problem this was automatically and exactly the case. However, here, since we did not consider the $S$ -operators neither in the Hamiltonian nor in the RPA operator, spin symmetry is violated. On the other hand this permits to evaluated the importance of the $S$ -operators. Normally the eigenvectors of the RPA matrix should be such that for charge ($ch$) excitations the operators $~J^{+}_{ph\uparrow} + J^{+}_{ph\downarrow}~$ and $~J^{-}_{ph\uparrow} + J^{-}_{ph\downarrow}~$ can be factored whereas for spin ($sp$) excitations the combinations $~J^{+}_{ph\uparrow} - J^{+}_{ph\downarrow}~$ and $~J^{-}_{ph\uparrow} - J^{-}_{ph\downarrow}~$ hold. Because of our violation of spin symmetry this factorisation is not exact. To have a measure of this violation we plot in Fig. \ref{figure8} the ratio
\be
r=\frac{|\X^{\nu}_{ph\uparrow}|-|\X^{\nu}_{ph\downarrow}|}{|\X^{\nu}_{ph\uparrow}|+|\X^{\nu}_{ph\downarrow}|}\label{rapp_amplitudes-ud}
\ee
For exact spin symmetry $r$ should be zero. From Fig. \ref{figure8} we see that the violation is on the level of a fraction of one percent. This, therefore justifies, a posteriori having neglected the scattering terms ($S$ -terms) in the Hamiltonian and RPA operator. A further indication that $S$ -terms are not important comes from the energy weighted sum rule. We know that the sum rule including the $S$ -terms is fullfilled in SCRPA \cite{13,18}. However, neglecting them gives a slight violation. Considering the exact relation
\be
L=R
\ee
with
\sba
\label{SRWE}
\ba
L &=& \sum\limits_\nu \left (E_\nu - E_0\right )\left|\langle \nu\left|F\right|0\rangle \right|^2 \nonumber \\
&=& \sum\limits_{\nu,|q|} \left (E_\nu - E_0 \right ) |\langle 0 |Q_{|q|,\nu}\,F|0\rangle |^2 ~~\label{Left} \\
&=& \sum\limits_{\nu,|q|} \left (E_\nu - E_0 \right ) |\langle 0 |\left[Q_{|q|,\nu} ,\,F\right]|0\rangle |^2 ~\nonumber \\
&=& \sum\limits_{\nu,|q|} \left (E_\nu - E_0 \right ) |\sum\limits_{i(|q|)}\sqrt{1-M_{i}}\left( \X^{\nu}_{i} + \Y^{\nu}_{i} \right)|^2
\nonumber \\
R &=& \frac{1}{2} \langle 0\left|\left[F, \left[H, F\right]\right]\right|0\rangle \nonumber \\
&=& \sum\limits_{i(|q|)} \sqrt{1-M_{i}} \sum\limits_{i'(|q|)}\sqrt{1-M_{i'}}\left( \A_{i, i'} - \B_{i,i'} \right)~~~
\label{Right}
\ea
\sea
with
\be
F = \sum\limits_{i (|q|)} \; \left( J^{+}_{i} + h.c\right)~,\label{op_Fgene6sit}
\ee
we trace in Fig. \ref{figure9} the ratio $\xi=\frac{R-L}{R}$. Again we see that the violation is on the level of a fraction of one percent, confirming the very small influence of the scattering terms.

\begin{figure}[ht]
  \includegraphics[width=8cm,height=7cm]{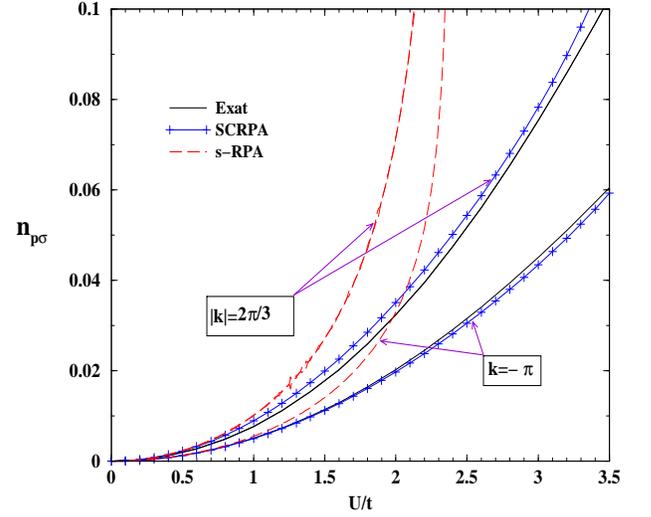}
      \caption{\label{figure10} Occupation numbers as function of the interaction $U$ for various values of the momenta $k$ for states above the Fermi level. For each approximation, s-RPA and SCRPA, the occupations numbers are represented in increasing order like $k$ ($-\pi $, $-\frac{2\pi }{3}$, $\frac{2\pi }{3}$). Let us notice that the modes $k=\frac{2\pi }{3}$ and $k=-\frac{2\pi }{3}$ are degenerate.}
\end{figure}

\begin{figure}[ht]
  \includegraphics[width=8cm,height=7cm]{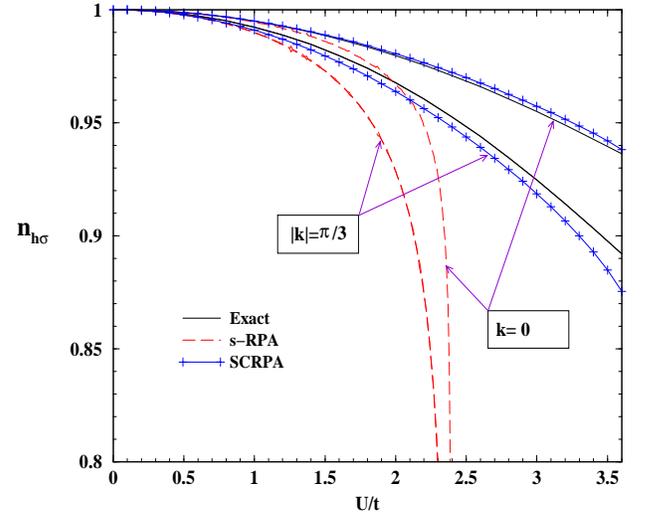}
      \caption{\label{figure11} Occupation numbers as function of the interaction $U$ for various values of the momenta $k$ for the holes states. For each approximation, s-RPA and SCRPA and exact solution, the occupation numbers are represented like $k=$ $0 $, $\frac{\pi}{3}$, $-\frac{\pi}{3}$. Let us notice that the modes $k=$ $\frac{\pi}{3}$ and $k=-\frac{\pi}{3}$ are degenerate.}
\end{figure}

A further quantity which crucially tests the ground state correlations are the occupation numbers. We have no direct access to them, however, we will use the so-called Catara approximation for their evaluation \cite{20} :
\sba
\label{nkb}
\ba
n_{p\sigma}=\langle \hat{n}_{p\sigma}\rangle &=& \sum\limits_{h} \langle  J^{+}_{ph,\sigma} \,J^{-}_{ph,\sigma}\rangle
\nonumber \\
&=& \sum\limits_{h} \left (1- \langle  M_{ph\sigma}\rangle\right)\sum\limits_{\nu} \left|\Y^{\nu}_{ph\sigma}\right|^2~,\label{np}
\\
n_{h\sigma}=\langle \hat{n}_{h\sigma}\rangle &=& \sum\limits_{p} \langle  J^{+}_{ph,\sigma} \,J^{-}_{ph,\sigma}\rangle
\nonumber \\
&=& 1- \sum\limits_{p} \left (1-\langle  M_{ph\sigma}\rangle \right)\sum\limits_{\nu} \left|\Y^{\nu}_{ph\sigma}\right|^2~.~~~ \label{nh}
\ea
\sea
We show these quantities in Figs. \ref{figure10} and \ref{figure11} in comparison with the exact values and the ones of standard RPA. We again see the excellent performance of SCRPA.

Concluding this section we can say that the expectation we had from the 2 -sites case, with its exact solution, have very satisfactorily also been fullfilled in the 6 -sites case. However, in spite of the very good performance of SCRPA, there is the limitation that SCRPA, in the symmetry conserving basis of plane waves used here, can not be employed in the strong $U$ limit. One also may wonder how the extension to cases with sites number $2+4n$ with $n>1$ works. For such cases it does not make sense any more to elaborate the Hamiltonian in its detailed form as given in eq.(\ref{ham_quasip6sita}). This explicit expression was only given to make clear the detailed internal structure of the approach for a definite example. In the general case with many sites one would just take the form (\ref{hamiltonian_imp}) of the Hamiltonian, calculate the double commutators as needed in (\ref{elements-of-RPA}) and then express the resulting correlation functions by the $\X$- and $\Y$- amplitudes. That such a program is feasable in terms of analytic work and numerical execution was demonstrated in our earlier work on the multilevel pairing model \cite{10} where cases up to hundred levels were treated. However, this number was not considered of an upper limit. Though the present model is slightly more complicated, we think that a generalisation to the case of many sites is perfectly possible. It needs, however, some investment which is planned for the future. This also concerns the $D=2$ case. Another question to ask is whether the degradation of the SCRPA results going from the $N=2$ to the $N=6$ case does not go on considering $N=10,~14,~$ etc? One again may cite the experience with the multilevel pairing model \cite{10} where also the $N=2$ case turned out to be exact in SCRPA but not the other cases. However, all $N>2$ cases showed more or less the same degrees of accuracy: excellent results of SCRPA up to the phase transtion point and deterioration beyond. Since this behavior has also been found in simpler models \cite{12} we think that this is a generic feature of SCRPA and that this behavior will also translate to the case of the present model.

Another problem for further work is how to continue the present theory into the strong coupling regime. Of course, there exists the possiblilty to perform SCRPA in the symmetry broken basis, but details and how to match with the summetry unbroken phase must still be worked out. Also the inclusion of higher order operators, as will shortly be discussed in the next section, may be an intersting direction  in this respect. 

\section{\label{4siteprob} Four sites problem}

\subsection{\label{4siteprob-unbrok-basis} The symmetry unbroken case}

\begin{figure}[ht]
  \includegraphics[width=6cm,height=3cm]{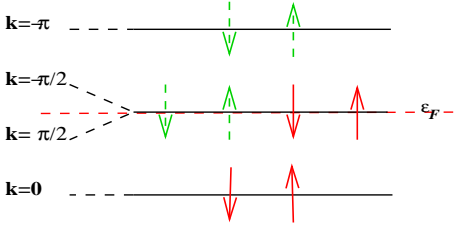}
      \caption{\label{figure12} Level spectrum for $U=0$ for the half filled chain with four sites with spin projection $m_{s} = 0$. The occupied states are represented by the full arrows and those not occupied are represented by the dashed arrows.}
\end{figure}

\begin{figure}[ht]
  \includegraphics[width=8cm,height=7cm]{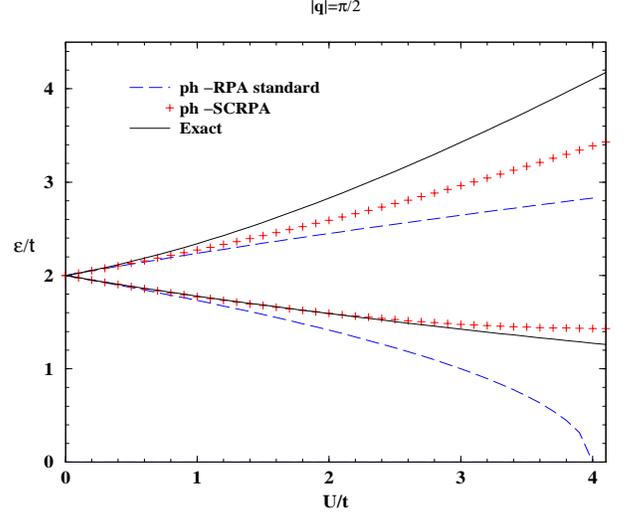}
      \caption{\label{figure13} Energies of excited states with standard RPA, SCRPA, and exact solution for four sites with spin projection $m_{s} = 0$ and for $|q|=\frac{\pi}{2}$ in the symmetry unbroken basis.}
\end{figure}

The problem of the 4 -sites case is easily located in regarding the level scheme of Fig. \ref{figure12} (see also ref \cite{21} deeling with the attractive Hubbard model in $1D$). 
We see that the Fermi energy coincides with the second level which is half filled. The uncorrelated groundstate is therefore degenerate and excitations with momentum transfer $|q|=\pi $ cost no energy. On the other hand for excitations with $|q|=\frac{\pi}{2} $ there is no problem. The corresponding RPA operator is given by 
\ba
Q^{\dagger}_{|q|=\frac{\pi}{2}, \nu}=& &
\X^{\nu}_{13,\uparrow}\, K^{+}_{31,\uparrow} +\X^{\nu}_{24,\uparrow}\,K^{+}_{42,\uparrow}
\nonumber \\
&+&\X^{\nu}_{13,\downarrow}\, K^{+}_{31,\downarrow} +\X^{\nu}_{24,\downarrow}\,K^{+}_{42,\downarrow}
\nonumber \\
&-&\Y^{\nu}_{13,\uparrow}\, K^{-}_{13,\uparrow}- \Y^{\nu}_{24,\uparrow}\,K^{-}_{24,\uparrow}
\nonumber \\
&-&\Y^{\nu}_{13,\downarrow}\,K^{-}_{13,\downarrow}-\Y^{\nu}_{24,\downarrow}\, K^{-}_{24,\downarrow}
\label{op-dexqpispi2}
\ea
In Fig. \ref{figure13} we show the results of s-RPA and SCRPA, together with the exact solution. We see that the lower excitation is still very well reproduced by SCRPA, whereas for the second excited state SCRPA only reduces the difference of s-RPA to exact by half. The real problem shows up for the transfer $|q|=\pi $. The corresponding operator is 
\ba
Q^{\dagger}_{|q|=\pi, \nu} =& &
 \X^{\nu}_{14,\uparrow}\, K^{+}_{41,\uparrow} +\X^{\nu}_{14,\downarrow}\,K^{+}_{41,\downarrow}
\nonumber \\
&+&\X^{\nu}_{23,\uparrow}\, K^{+}_{32,\uparrow} + \X^{\nu}_{23,\downarrow}\,K^{+}_{32,\downarrow}
\nonumber \\
&-&\Y^{\nu}_{14,\uparrow}\, K^{-}_{14,\uparrow} - \Y^{\nu}_{14,\downarrow}\, K^{-}_{14,\downarrow}
\nonumber \\
&-&\Y^{\nu}_{23,\uparrow}\, K^{-}_{23,\uparrow} - \Y^{\nu}_{23,\downarrow}\, K^{-}_{23,\downarrow}
\label{op-dexpi}
\ea
\begin{figure}[ht]
  \includegraphics[width=8cm,height=7cm]{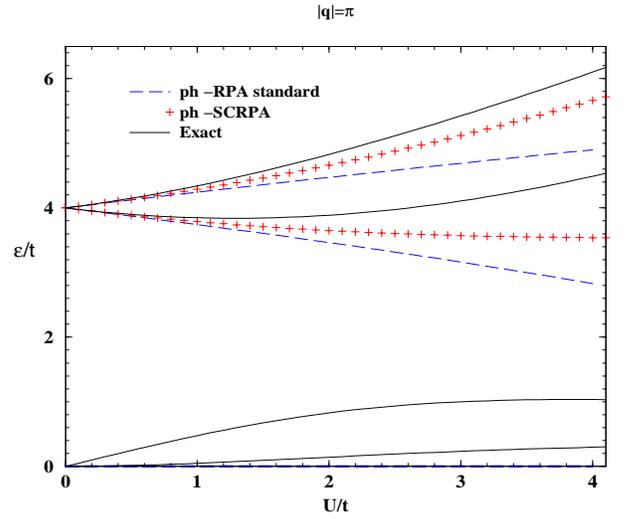}
      \caption{\label{figure14} Energies of excited states with standard RPA, SCRPA, and exact solution for four sites with spin projection $m_{s} = 0$ and for $|q|=\pi$ in the symmetry unbroken basis.}
\end{figure}
The standard RPA produces a doubly degenerate zero mode independent of $U$ as seen on Fig. \ref{figure14}. As compared with the exact solution, we see that these two zero modes approximate two very low lying exact solutions. Unfortunately, because of these modes at low energy the SCRPA could not be stabilised. The only possibility consisted in excluding the components $K^{\pm}_{32,\uparrow} $ and  $K^{\pm}_{32,\downarrow}$ in the RPA operator. Then the self consistency was achieved without problem and the result is shown in Fig. \ref{figure14}. The result of SCRPA is half way in between s-RPA and the exact solution. On the other hand, because of the omission of the two lower states, the groundstate energy can not correctly be calculated in SCRPA. Therefore, for the 4 -sites problem in the symmetry unbroken basis (plane waves), the SCRPA cannot fully account for the situation.

\subsection{\label{4siteprob-brok-basis} Symmetry broken basis}

\begin{figure}[ht]
  \includegraphics[width=8cm,height=7cm]{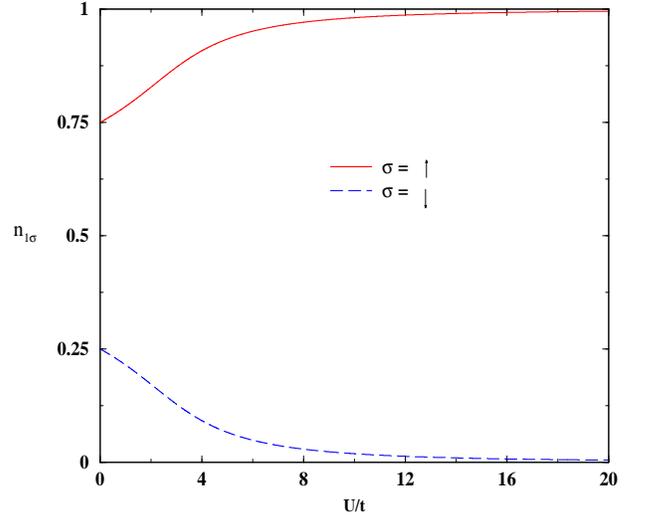}
      \caption{\label{figure15} Occupation numbers for site $1$, $n_{1,\uparrow}$ et $n_{1,\downarrow}$, as a function of interaction $U$ in the symmetry broken basis.}
\end{figure}

An analysis of the HF solution shows that, as soon as $U\neq 0$, the plane wave state becomes unstable and the system prefers a staggered magnetisation. The general HF transformation can be written as
\sba
\be
\left(\barr{c} c^{\dagger}_{1,\uparrow} \\ c^{\dagger}_{2,\uparrow}  \\
c^{\dagger}_{3,\uparrow} \\ c^{\dagger}_{4,\uparrow} \earr \right)
=  \frac 1{\sqrt{2}} \left(\barr{cccc}
v & -1 &  0 &  u \\
u &  0 & -1 & -v \\ 
v &  1 &  0 &  u \\ 
v &  0 &  1 & -v \earr \right)
\left(\barr{c}  a^{\dagger}_{1,\uparrow} \cr a^{\dagger}_{2,\uparrow} \cr 
a^{\dagger}_{3,\uparrow} \cr a^{\dagger}_{4,\uparrow} \earr \right)\mbox{,}
\label{vecppuinv_hf4}
\ee
\be
\left(\barr{c} c^{\dagger}_{4,\downarrow} \cr c^{\dagger}_{3,\downarrow}  \cr
c^{\dagger}_{2,\downarrow} \cr c^{\dagger}_{1,\downarrow} \earr \right)
=  \frac 1{\sqrt{2}} \left(\barr{cccc}
v & -1 &  0 &  u \cr 
u &  0 & -1 & -v \cr
v &  1 &  0 &  u \cr 
v &  0 &  1 & -v \cr \earr \right) 
\left(\barr{c} a^{\dagger}_{1,\downarrow} \cr a^{\dagger}_{2,\downarrow} \cr
a^{\dagger}_{3,\downarrow} \cr a^{\dagger}_{4,\downarrow} \earr \right)\mbox{,}
\label{vecppdinv_hf4inv}
\ee
\sea
with $u=\cos(\vartheta )$ and $v=\sin(\vartheta )\,e^{i\,\varphi}$. The minimisation of the groundstate energy with
\be
| HF\rangle = a^{\dagger}_{1,\uparrow} a^{\dagger}_{1,\downarrow}a^{\dagger}_{2,\uparrow}
a^{\dagger}_{2,\downarrow} |-\rangle \mbox{,}
\label{state_hf4}
\ee 
shows that $\varphi =0$ for any value of $U$ and the angle $\vartheta $ is obtained from
\be
\tan^4(\vartheta) -\frac U{2\,t}\,\tan^3(\vartheta) - 1 = 0~\mbox{.}
\ee
The occupation numbers are given by 
\ba
n_{1,\uparrow} & = & n_{3,\uparrow} = n_{2,\downarrow} = n_{4,\downarrow}
= \frac 1{2} \left ( 1 + sin^2(\vartheta)\right )
\nonumber \\
n_{1,\downarrow} & = & n_{3,\downarrow} = n_{2,\uparrow} = n_{4,\uparrow}
= \frac 1{2} \, cos^2(\vartheta)
\label{densHF_4sites}
\ea
and shown in Fig. \ref{figure15} which illustrates the spontaneous symmetry breaking for any value of $U$. For $U\rightarrow \infty $ we have a perfect antiferromagnet.

\begin{figure}[ht]
     \includegraphics[width=8cm,height=7cm]{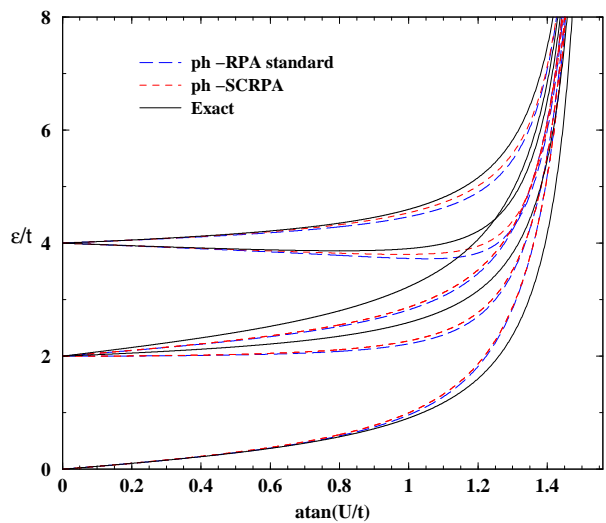}
      \caption{\label{figure16} Energies of excited states with standard RPA, SCRPA, and exact solution as a function of $U$ for four sites with spin projection $m_{s} = 0$ in the symmetry broken basis.}
\end{figure}

\begin{figure}[ht]
  \includegraphics[width=8cm,height=7cm]{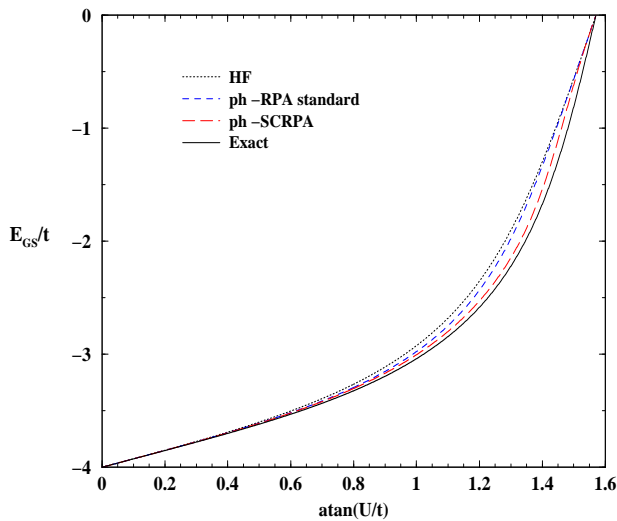}
      \caption{\label{figure17} Groundstate energies in HF, standard RPA, SCRPA, and exact solution as a function of $atan(\frac{U}{t})$ for four sites with spin projection $m_{s} = 0$ in the symmetry broken basis.}
\end{figure}

We can now perform a SCRPA calculation in the symmetry broken basis. The RPA operators are given by
\ba
Q^{\dagger}_{\sigma \nu} =&& \X^{\nu}_{1\sigma, 3\sigma}\, K^{+}_{3\sigma,1\sigma} +\X^{\nu}_{2-\sigma, 4-\sigma}\,K^{+}_{4-\sigma, 2-\sigma} 
\nonumber \\
&-& \Y^{\nu}_{1\sigma, 3\sigma}\, K^{-}_{1\sigma, 3\sigma}-\Y^{\nu}_{2-\sigma, 4-\sigma}\,K^{-}_{2-\sigma, 4-\sigma}
\label{op-dexqpis1}
\ea
with $\sigma=\pm\frac 1 2$. We also have two other excitation operators
\ba
Q^{\dagger}_{1\nu} =&& \X^{\nu}_{1\uparrow, 4\uparrow}\, K^{+}_{4\uparrow,1\uparrow} +
\X^{\nu}_{1\downarrow, 4\downarrow}\,K^{+}_{4\downarrow, 1\downarrow}
\nonumber \\
&-& \Y^{\nu}_{1\uparrow, 4\uparrow}\, K^{-}_{1\uparrow, 4\uparrow}-
\Y^{\nu}_{1\downarrow, 4\downarrow}\,K^{-}_{1\downarrow, 4\downarrow}
\label{op-dexqpis11}
\ea
and
\ba
Q^{\dagger}_{2\nu} =&& \X^{\nu}_{2\uparrow, 3\uparrow}\, K^{+}_{3\uparrow,2\uparrow} +
\X^{\nu}_{2\downarrow, 3\downarrow}\,K^{+}_{3\downarrow, 2\downarrow}
\nonumber \\
&-&\Y^{\nu}_{2\uparrow, 3\uparrow}\, K^{-}_{2\uparrow, 3\uparrow}-
\Y^{\nu}_{2\downarrow, 3\downarrow}\,K^{-}_{2\downarrow, 3\downarrow}
\label{op-dexqpis12}
\ea
In Figs. \ref{figure16} and \ref{figure17} we give the results. The most striking feature is that s-RPA and SCRPA are very close and that the error with respect to the exact solution does not become greater than $25\%$ for any value of $U$. Though the improvement of SCRPA over s-RPA is very small in each channel, at the end in the groundstate energy this sums to a more substantial correction in the right direction for the ground state energy. This is shown in Fig. \ref{figure17} as a function of $atan(\frac{U}{t})$. We see that HF, s-RPA and SCRPA become exact for $U=0$ and $U\rightarrow \infty $. In between SCRPA deviates e.g. by $8\%$ from the exact result at $U\simeq 6$ ($atan(\frac{U}{t}) \simeq 1.4$) whereas this deviation is $20\%$ for s-RPA.

Concluding this section on the 4 -sites case at half filling we can say that in the symmetry unbroken basis SCRPA is unable to account for some low lying excitations and therefore fails to reproduce the groundstate energy as well. In the symmetry broken basis SCRPA gives very little correction over s-RPA. However, the maximum error is not greater than $25\%$ for all values of $U$ for the excited states and the groundstate energy in SCRPA whereas this is $30\%$ for standard RPA. This may be an interesting result in view of the importance of the so-called `plaquettes' (see e.g. ref \cite{22}) in high $T_c$ superconductivity. Nevertheless, even though one plaquette (4 -sites) may reasonably be described, the present approach can not account for the situation of many plaquettes in interaction which is the real situation in 2D. For the future it is therefore very interesting to develope an extension of the present SCRPA which not only gives the exact solution for the 2 -sites case but equally for the 4 -sites case. Such a generalisation is possible in including into the RPA operator in addition to the Fermion pair operators also quadruples of Fermion operators. This is a general principle and it has already been demonstrated to hold true in the case of the simpler Lipkin model \cite{23}. One could call such an extension a second SCRPA in analogy to the well known standard second RPA which involves in addition to the $ph$ configurations also $2p-2h$ ones. In the case of many plaquettes this second SCRPA would then constitute a self consistent mean field theory for plaquettes.

\section{\label{Conclusions} Discussion, conclusions and outlook}

In this work a many body approach which has essentially been developed in the nuclear physics context in recent years \cite{9} has been applied to the Hubbard model for finite number of sites. The theory is an extension of standard RPA, called Self Consistent RPA (SCRPA), which aims to correct its well known deficiencies as the quasiboson approximation with its ensuing violation of the Pauli principle and its perturbation theoretical aspect. Of course the appealing features of RPA, as for instance fullfillment of sum rules, restoration of broken symmetries, Goldstone theorem, numerical practicability and physical transparency should be kept as much as possible. That this is indeed the case with SCRPA has in the past been demonstrated with applications to several non trivial models \cite{10} as for instance the many level pairing (Richardson) model \cite{10} and the 3-level Lipkin model \cite{11}. SCRPA can be derived by minimising an energy weighted sum rule and it is therefore a non perturbative variational approach though it is in general not of the Raleigh Ritz type. The resulting equations are a non linear version of the RPA type which can be interpreted as the mean field equations of interacting quantum fluctuations. Though the SCRPA equations are of the Schr\"odinger type, their non linearity non the less makes their numerical solution quite demanding. We therefore thought it indicated to begin with applications to the Hubbard model restricting them to low dimensional cases given by a finite number of sites where exact diagonalisation can easily be obtained. We then logically started out considering the two sites case (with periodic boundary conditions), increasing the number of sites by steps of two, i.e $N=2,4,6, \ldots $. To our satisfaction SCRPA solves the 2 -sites problem exactly for any value of $U$. This, as a matter of fact, did not come entirely as a surprise, since the same happened already with the pairing problem for two fermions \cite{10} and indeed it can be shown that SCRPA solves a general two body problem exactly \cite{24}. It is nontheless worth pointing out that other respectable many body theories fail in the two particle case, apart from the low $U$--limit.

In the four sites problem at half filling SCRPA failed. This, as in all $4n$ ($n=1,2,3, \ldots $) cases, presents the particular problem that the system is unstable with respect to the formation of staggered magnetisation for any finite value of $U$ and this prevented the SCRPA solution to exist in the plane wave basis for particular values of the momentum transfer $|q|$. At the end of the paper we indicated that extending the present RPA ansatz of $ph$ pairs to include quadruples of fermion operators can solve not only the two electrons but also the four electrons case exactly. This is particularly interesting in view of the fact that the 4 -sites case (plaquette) may be very important for the explanation of high $T_c$ superconductivity, in considering the many plaquette configurations in 2D \cite{22}. In this work we jumped directly to the six sites problem which, as all $2+4n$ cases, causes no particular difficulties in SCRPA, even in the symmetry unbroken basis of plane waves. Of course, in the case of 6 -sites, SCRPA is not exact any more. However, it is shown that the results are still excellent for all quantities considered : excited states, groundstate, and occupation numbers. Contrary to the two sites case, the SCRPA solutions in the plane wave basis cannot be obtained for all values of $U$. Somewhere after the point where, as a function of $U$, the first mean field instability shows up, the SCRPA also starts to deteriorate and in fact does not converge any longer. Often the mean field critical value of $U$ is by passed by $20\%$ up to $50\%$ in SCRPA, still staying excellent. However, to go into the strong $U$--limit we have to introduce the above mentioned quadruple fermion operators or perform a SCRPA calculation in the symmetry broken basis \cite{12}. Such investigations shall be left for the future. We also gave arguments why we think that, going to the $N>6$ cases, the precision we found for $N=6$ will not deteriorate. We therfore think that our formalism will allow to find precise results for system sizes where an exact diagonalisation becomes prohibitive. Problems in 2D with closed shell configurations probably also can and shall be considered with the present formalism. Also, as shown in \cite{10}, the extension to finite temperatures is possible.

We also should mention that in this work we neglected the so-called scattering terms of the form $a^{\dagger}_{p}\,a_{p'}$ or $a^{\dagger}_{h}\,a_{h'}$, that is fermion $ph$ operators where either both indices are above or both below the Fermi level. In standard RPA those configurations automatically decouple from the $ph$ and $hp$ spaces. However, in SCRPA with its rounded occupation numbers, there is formally no reason not to include them. As a matter of fact, as shown in earlier work \cite{11,18}, to assure the fullfillment of the f- sum rule and the restoration of broken symmetries, these scattering terms must be taken into account. In the present case, as well as in earlier studies, the scattering terms seem to be almost linearly  dependent with the ordinary $ph$ and $hp$ configurations. This fact induced difficulties with the iteration procedure, since they correspond to very small eigenvalues of the norm matrix. Though we do not exclude the possibility that this difficulty could be mastered with a more refined numerical algorithm, we finally refrained from persuing this effort, since we could show that the influence of the scattering terms on the results is only on the level of a fraction of percent and also the f- sum rule is only violated on this order.

In short we showed that SCRPA, as in previous models, performs excellently in the symmetry unbroken regime of the Hubbard model. However, the high $U$--limit and the $4n$ sites cases need further developements.

\begin{acknowledgments}
We are very grateful to B. K. Chakraverty and J. Ranninger for elucidating discussions. One of us (P.S.) thanks A. M. Tremblay for useful information. One of the authors (J.D.) acknowledges the support from the Spanish DGI under grant BFM2003-05316-C02-02.
\end{acknowledgments}
\begin{appendix}

\section{ particle--hole correlation Functions }
\label{correl-ph}

We give the commutations rules which will be useful in the calculation of the correlations functions in the $ph$ channel,
\ba
\left[ Q_\nu,Q^\dagger_{\nu'}\right] &=& \sum\limits_i \left( \X^\nu_i \X^{\nu'}_i - \Y^\nu_i \Y^{\nu'}_i \right) \frac {1 - M_i} {1 - \langle M_i \rangle}~,\nonumber\\
\left[ Q_\nu,Q_{\nu'}\right] &=& \sum\limits_i \left( \Y^\nu_i \X^{\nu'}_i - \X^\nu_i \Y^{\nu'}_i \right) \frac {1 - M_i} {1 - \langle M_i \rangle } ~,\\
\left[ M_i, Q_\nu \right] &=& - 2 \Y^\nu_i \sum\limits_{\nu_1}\left( \X^{\nu_1}_i Q^\dagger_{\nu_1} + \Y^{\nu_1}_i Q_{\nu_1} \right) \nonumber\\
\left[ M_i, Q^\dagger_\nu \right] &=& 2 \Y^\nu_i \sum\limits_{\nu_1}\left( \Y^{\nu_1}_i Q^\dagger_{\nu_1} + \X^{\nu_1}_i Q_{\nu_1} \right)~. \nonumber
\ea
Thus, the following average values can be calculated (commuting the $Q$'s to the right)

\ba
\langle Q_{\nu_3} Q^\dagger_{\nu_2} Q_{\nu_1} Q^\dagger_{\nu_0} \rangle = \sum\limits_{i j} 
\frac { (\X^{\nu_3}_{i} \X^{\nu_2}_{i}-\Y^{\nu_3}_{i} \Y^{\nu_2}_{i})} {(1 - \langle M_{i} \rangle)}
\nonumber \\
.\frac {(\X^{\nu_1}_{j} \X^{\nu_0}_{j} -\Y^{\nu_1}_{j} \Y^{\nu_0}_{j})}
{(1 - \langle M_{j} \rangle)} \langle (1 -M_{i}) (1-M_{j}) \rangle 
\ea
\ba
\langle Q_{\nu_3} \left[ Q_{\nu_1}, Q^\dagger_{\nu_2} \right] Q^\dagger_{\nu_0} \rangle =
\sum\limits_{i j} 
\frac { (\X^{\nu_3}_{i} \X^{\nu_0}_{i} - \Y^{\nu_3}_{i} \Y^{\nu_0}_{i})} {(1 - \langle M_{i} \rangle)} 
\nonumber \\
.\frac {(\X^{\nu_1}_{j} \X^{\nu_2}_{j} - \Y^{\nu_1}_{j} \Y^{\nu_2}_{j})}
{(1 - \langle M_{j} \rangle)} \langle (1 -M_{i}) (1-M_{j}) \rangle 
\nonumber \\
-2 \sum_{i } {\frac {\X^{\nu_3}_{i} \X^{\nu_2}_{i} \X^{\nu_1}_{i} \X^{\nu_0}_{i} - \Y^{\nu_3}_{i} \Y^{\nu_2}_{i} \Y^{\nu_1}_{i} \Y^{\nu_0}_{i}} {(1 - \langle M_{i} \rangle)}}
\ea
Finally, one can express the correlation function according to the amplitudes RPA, $\langle M_i \rangle$ and of $\langle M_i M_j\rangle$ as
\ba
\langle Q_{\nu_3} Q_{\nu_1} Q^\dagger_{\nu_2} Q^\dagger_{\nu_0} \rangle =
\langle Q_{\nu_3} \left[ Q_{\nu_1}, Q^\dagger_{\nu_2} \right] Q^\dagger_{\nu_0} \rangle 
\nonumber \\
+\langle Q_{\nu_3} Q^\dagger_{\nu_2} Q_{\nu_1} Q^\dagger_{\nu_0} \rangle
\nonumber\\
\nonumber\\
= 2 \sum\limits_{i j} 
\frac { (\X^{\nu_3}_{i} \X^{\nu_2}_{i}-\Y^{\nu_3}_{i} \Y^{\nu_2}_{i})} {(1 - \langle M_{i} \rangle)} 
\nonumber \\
.\frac {(\X^{\nu_1}_{j} \X^{\nu_0}_{j} -\Y^{\nu_1}_{j} \Y^{\nu_0}_{j})}
{(1 - \langle M_{j} \rangle)} \langle (1 -M_{i}) (1-M_{j}) \rangle 
\nonumber \\
+\sum\limits_{i j} \frac { (\X^{\nu_3}_{i} \X^{\nu_0}_{i} - \Y^{\nu_3}_{i} \Y^{\nu_0}_{i})} {(1 - \langle M_{i} \rangle)} 
\nonumber\\
.\frac {(\X^{\nu_1}_{j} \X^{\nu_2}_{j} - \Y^{\nu_1}_{j} \Y^{\nu_2}_{j})}
{(1 - \langle M_{j} \rangle)} \langle (1 -M_{i}) (1-M_{j}) \rangle 
\nonumber \\
-2 \sum_{i } \frac {\X^{\nu_3}_{i} \X^{\nu_2}_{i} \X^{\nu_1}_{i} \X^{\nu_0}_{i} - \Y^{\nu_3}_{i} \Y^{\nu_2}_{i} \Y^{\nu_1}_{i} \Y^{\nu_0}_{i}} {(1 - \langle M_{i} \rangle)}
\nonumber \\
\ea

\section{Density-density correlation Functions }
\label{dens_quasipart}

Given that this RPA formalism preserves the number of particles per spin -$\sigma$ (owing to the fact that the transformation HF does not break the symmetry of spin), one has
\be
\hat{N}_{\sigma} = N_{\sigma} + \sum\limits_{p} \tilde{n}_{p\sigma} -\sum\limits_{h} \tilde{n}_{h\sigma}
\ee
and the average value $\langle \hat{N}_{\sigma}\rangle = N_{\sigma} = \frac N{2}$ what gives us
\be
\sum\limits_{p} \langle \tilde{n}_{p\sigma}\rangle = \sum\limits_{h} \langle \tilde{n}_{h\sigma}\rangle
\label{np=nh}
\ee
On the other hand, one also has
\ba
\hat{N}_{\sigma}\hat{N}_{\sigma'} =&&(N_{\sigma} + \sum\limits_{p} \tilde{n}_{p\sigma} -\sum\limits_{h} \tilde{n}_{h\sigma} ) 
\nonumber \\
&&~~~~~.(N_{\sigma'} + \sum\limits_{p'} \tilde{n}_{p'\sigma'} -\sum\limits_{h'} \tilde{n}_{h'\sigma'} )
\ea
with the average value $\langle \hat{N}_{\sigma}\hat{N}_{\sigma'}\rangle = N_{\sigma} + N_{\sigma'}$, which gives us
\ba
\langle (\sum\limits_{p} \tilde{n}_{p\sigma} -\sum\limits_{h} \tilde{n}_{h\sigma} ) (\sum\limits_{p'} \tilde{n}_{p'\sigma'} -\sum\limits_{h'} \tilde{n}_{h'\sigma'} )\rangle = 
\nonumber \\
N_{\sigma'}\langle (\sum\limits_{p} \tilde{n}_{p\sigma} -\sum\limits_{h} \tilde{n}_{h\sigma} )\rangle
\nonumber \\
+N_{\sigma} \langle (\sum\limits_{p'} \tilde{n}_{p'\sigma'} -\sum\limits_{h'} \tilde{n}_{h'\sigma'} )\rangle
\label{N2}
\ea
Thus for our case, there is the relation
\ba
\langle (\sum\limits_{p} \tilde{n}_{p\uparrow} -\sum\limits_{h} \tilde{n}_{h\uparrow} )(\sum\limits_{p'} \tilde{n}_{p'\downarrow} - \sum\limits_{h'} \tilde{n}_{h'\downarrow} )\rangle = 
\nonumber \\
3 (\sum\limits_{p\sigma} \langle\tilde{n}_{p\sigma} \rangle  -\sum\limits_{h\sigma} \langle\tilde{n}_{h\sigma} \rangle )= 0
\label{nni=nni}
\ea
\end{appendix}



\begin{references}

\bibitem{1} A.L.~Fetter, J.D.~Walecka, ~~Quantum Theory of Many-Particle Systems, ~~Mc\-Graw-Hill, Inc. ~~1971.
\bibitem{2} G.D.~Mahan, ~~Many-Particle Physics, ~~Plenum Press, New York, ~~1981.
\bibitem{3} F. Furche, Phys. Rev. {\bf B 64}, (2001) 195120
\bibitem{4} G. F. Bertsch, C. Guet, K. Hagino, physics/0306058
\bibitem{5} A. K. Kerman and C. Y. Lin, Ann. of Phys. (N.Y.) {\bf 241} (1995) 185;\\
            A. K. Kerman and C. Y. Lin, Ann. of Phys. (N.Y.) {\bf 269} (1998) 55.
\bibitem{6} P.~Ring, P.~Schuck, {\em The Nuclear Many--Body Problem}, Springer, Berlin 1980
\bibitem{7} G. Baym, L. P. Kadanoff, Phys. Rev  {\bf 124} (1961) 28; \\
            G. Baym, Phys. Rev  {\bf 127} (1962) 1391; \\
            G. Baym, Phys. Lett.  {\bf 1} (1962) 242.
\bibitem{8} J.P.~Blaizot, G.~Ripka, {\em Quantum Theory of Finite Systems}, The MIT Press, Cambridge, 1986.
\bibitem{9} P. Schuck, S. Ethofer, Nucl. Phys. {\bf A 212} (1973) 269; \\
            J. Dukelsky and P. Schuck, Nucl. Phys. {\bf A 512} (1990) 466; \\
	    J. Dukelsky, G. R\"opke, P. Schuck, Nucl. Phys. {\bf A 628} (1998) 17.
\bibitem{10} P. Kr\"uger, P. Schuck, Europhy. Lett. {\bf 27} (1994) 395;\\
             J. G. Hirsch, A. Mariano, J. Dukelsky, P. Schuck, Ann. Phys. (NY) 296 (2002) 187;\\
	     A. Storozhenko, P. Schuck, J. Dukelsky, G. R\"opke and A. Vdovin, Ann. Phys. (NY) {\bf 307} (2003) 308-334.
\bibitem{11} D. Delion, P. Schuck, J. Dukelsky, submitted in Phy. Rev. {\bf C} (nucl-th/0405002).
\bibitem{12} A. Rabhi, R. Bennaceur, G. Chanfray, P. Schuck, Phys. Rev. {\bf C 66} (2002) 064315.
\bibitem{13} D. S. Sch\"afer and P. Schuck, Phys. Rev. B {\bf 59}, (1999) 1712-1733.
\bibitem{14} F. Aryasetiawan, T. Miyake and K. Terakura, Phys. Rev. Lett. {\bf 88} (2002) 166401.
\bibitem{15} G.~Seibold, F.~Becca, J.~Lorenzana, Phys. Rev. B {\bf 67},  085108 (2003).
\bibitem{16} Y.M. Vilk, L.~Chen,  A.--M.S.~Tremblay, Phys. Rev. B {\bf 49}, 13\,267 (1994); {\em ibid.} Physica C {\bf 235-240}, 2235 (1994); \\
             Y.M. Vilk et A.--M.S.~Tremblay, J. Phys. I France {\bf 7} (1997) 1309-1368;\\
             S.~Allen and A.--M.S.~Tremblay, Phys. Rev. B {\bf 64},  075115 (2001);\\
             B.~Kyung, J. S. ~Landry, A.--M.S.~Tremblay, Phys. Rev. B {\bf 68},  174502 (2003).
\bibitem{24} D. Delion, P. Schuck, to be published.
\bibitem{17} H.M. Sommermann. Ann. Phys. 151 (1983), p. 163;\\
             D. Vautherin and N. Vinh Mau, Nucl. Phys. {\bf A 422} (1984) 140;\\
	     N. Vinh Mau and D. Vautherin, Nucl. Phys. {\bf A 445} (1985) 245.
\bibitem{18} M.~Grasso and F.~Catara, Phys. Rev. {\bf C 63}, 014317 (2000).
\bibitem{19} D. Janssen and P. Schuck, Z. Physik {\bf A 339} (1991) 34-50.
\bibitem{20} F.~Catara, G. Piccitto, M. Sambataro and N. Van Giai, Phys. Rev. {\bf B 54} (1996-II) 17536.
\bibitem{21} K. Tanaka and F. Marsiglio, Phys. Rev. {\bf B 60},  3508 (1999).
\bibitem{22} E. Altman, A. Auerbach, Phys. Rev. {\bf B 65}, 104508 (2002).
\bibitem{23} A. Storozhenko, P. Schuck, to be published.
\end{references}
\end{document}